\documentclass[twocolumn,epjc3]{svjour3}
\pdfoutput=1 

\usepackage{amsmath}
\usepackage{graphicx}
\usepackage{amssymb}
\usepackage{subfigure}
\usepackage{cancel}
\usepackage{xspace}
\usepackage{relsize}
\usepackage{bbold}
\usepackage{lineno}
\usepackage{braket}
\usepackage{slashed}
\usepackage{multirow}
\usepackage{placeins}
\usepackage{rotating}
\usepackage{url}
\usepackage{color}
\usepackage{hyperref}
\usepackage{cite}
\usepackage{multicol}
\RequirePackage{fix-cm}

\definecolor{oucrimsonred}{rgb}{0.6, 0.0, 0.0}
\definecolor{persianblue}{rgb}{0.11, 0.22, 0.73}
\definecolor{forestgreen}{rgb}{0.13,0.35,0.13}
\hypersetup{colorlinks, citecolor=oucrimsonred, linkcolor=persianblue, urlcolor=oucrimsonred}

\newcommand{\HEPfit}{\texttt{HEPfit}\xspace}

\usepackage[T1]{fontenc} 

\smartqed  
\RequirePackage{graphicx}
\journalname{Eur. Phys. J. C}
\begin{document}
\title{On Flavourful Easter eggs for New Physics hunger and Lepton Flavour Universality violation}
\author{Marco Ciuchini\thanksref{em1,addr1}
\and
Ant\'onio M. Coutinho\thanksref{em2,addr1,addr2}
\and
Marco Fedele\thanksref{em3,addr3,addr4}
\and
Enrico Franco\thanksref{em4,addr4}
\and
Ayan Paul\thanksref{em5,addr4}
\and
Luca Silvestrini\thanksref{em6,addr4}
\and
Mauro Valli\thanksref{em7,addr4}
}
%
\thankstext{em1}{marco.ciuchini@roma3.infn.it}
\thankstext{em2}{coutinho@fis.uniroma3.it}
\thankstext{em3}{marco.fedele@uniroma1.it}
\thankstext{em4}{enrico.franco@roma1.infn.it}
\thankstext{em5}{ayan.paul@roma1.infn.it,}
\thankstext{em6}{luca.silvestrini@roma1.infn.it}
\thankstext{em7}{mauro.valli@roma1.infn.it}
\institute{INFN, Sezione di Roma Tre, Via della Vasca Navale 84,
  I-00146 Roma, Italy \label{addr1}
\and
  Dipartimento di Matematica e Fisica,
  Universit\`a di Roma Tre, Via della Vasca Navale 84,
  I-00146 Roma, Italy \label{addr2}
\and
  Dipartimento di Fisica,
  Universit\`a di Roma ``La Sapienza'', P.le A. Moro 2, I-00185 Roma,
  Italy  \label{addr3}
\and
  INFN, Sezione di Roma, P.le A. Moro 2,
  I-00185 Roma, Italy \label{addr4}
}
\date{Received: date / Accepted: date}
\maketitle
\begin{abstract}{Within the standard approach of
  effective field theory of weak interactions for $\Delta B = 1$
  transitions, we look for possibly unexpected subtle New Physics
  effects, here dubbed ``flavourful Easter eggs''. We perform a Bayesian
  global fit using the publicly available \HEPfit package, taking into account state-of-the-art
  experimental information concerning these processes, including the
  suggestive measurements from LHCb of $R_{K}$ and $R_{K^{*}}$, the
  latter available only very recently. We parametrize New Physics
  contributions to $b \to s$ transitions in terms of shifts of Wilson
  coefficients of the electromagnetic dipole and semileptonic
  operators, assuming CP-conserving effects, but allowing in general
  for violation of lepton flavour universality.  We show how
  optimistic/conservative hadronic estimates can impact quantitatively
  the size of New Physics extracted from the fit. With a conservative
  approach to hadronic uncertainties we find nonzero New Physics contributions
  to Wilson coefficients at the level of $\sim 3\sigma$, depending on the model
  chosen. Furthermore, given the interplay between hadronic
  contributions and New Physics effects in the leptonic vector current, a
  scenario with nonstandard leptonic axial currents is comparable to
  the more widely advocated one with New Physics in the leptonic vector
  current.}
  \end{abstract}

\section{Introduction}
\label{sec:intro}

Easter eggs nowadays also refer to inside jokes and/or secret messages
usually hidden e.g. in computer gaming and hi-tech software. In this
work, we take advantage of this terminology to motivate the search for
New Physics Beyond the Standard Model in the radiative and in the
(semi)leptonic channels of rare $B$ meson decays.

In the decades that have
followed the original formulation of flavour mixing~\cite{Cabibbo:1963yz},
the flavour structure of the SM has been experimentally
tested and well established. The tremendous progress of the
experimental facilities has probed the flavour of the SM to an
exquisite level of precision \cite{Amhis:2016xyh}, along with the
substantial effort on the part of the theoretical community to go well
beyond leading order computations \cite{Buras:2011we}. From this
perspective of ``precision tests'', radiative and (semi)leptonic
$\Delta B = 1$ processes, related at the partonic level to
$b \to s \gamma, s \ell \ell$ transitions, occupy a special place in
probing the SM and its possible extensions in
terms of New Physics (NP)
models~\cite{Beaujean:2013soa,Blake:2016olu}.

Firstly, these rare $B$ meson decays belong to the class of
flavour-changing neutral current (FCNC) processes, that are well known
to be sensitive probes of Physics Beyond the Standard Model (BSM): in
fact -- within the SM -- the flavour structure of the theory allows
FCNC to arise only at loop level, as a consequence of
the GIM mechanism \cite{Glashow:1970gm}.  This allows
for significant room for heavy new degrees of freedom to sizably
contribute to these rare processes.

Secondly, from the experimental side, the study of rare $B$ meson
decays offers us some of the most precise measurements amongst the
$| \Delta F | = 1$ processes. For instance, the measurement of the
inclusive branching fraction of $B \to X_{s} \gamma$ is currently
performed with a relative uncertainty of a few percent
\cite{Saito:2014das,Belle:2016ufb,Lees:2012ym}, while the
study of an exclusive mode such as $B \to K^{*} \ell \ell$ allows for
a detailed analysis of the angular distribution of the four final
state particles, yielding rich experimental information in terms
of angular functions of the dilepton invariant mass, with full
kinematic coverage of the latter \cite{Aaij:2013iag} and -- starting
from ref.~\cite{LHCb:2015dla} -- also with available experimental
correlations among the angular observables.

In $B$ Physics, the recent years have been characterized by the
emergence of a striking pattern of anomalies in multiple independent
studies of some of these rare $b \to s$ transitions
\cite{Blake:2017wjz}. Of particular importance, the measurement of the
$P_{5}'$ angular observable
\cite{Matias:2012xw,DescotesGenon:2012zf,Descotes-Genon:2013vna,Matias:2014jua}
stands out from all the other ones related to the angular distribution
of $B \to K^{*} \mu \mu\,$; first realized by the LHCb collaboration
\cite{Aaij:2013qta,Aaij:2015oid} and later on also by the Belle
collaboration \cite{Abdesselam:2016llu}, the experimental analysis of
$P_{5}'$ in the large recoil region of the decay points to a deviation
of about $3\sigma$ with respect to the SM prediction presented in
ref.~\cite{Descotes-Genon:2014uoa}. The latter, however, suffers from
possible hadronic uncertainties which are sometimes even hard to
guesstimate
\cite{Jager:2012uw,Jager:2014rwa,Lyon:2014hpa,Ciuchini:2015qxb}, and
this observation has been at the origin of a quite vivid debate in the
recent literature about the size of (possibly) known and (yet) unknown
QCD power corrections to the amplitude of this process in the infinite
mass
limit~\cite{Hurth:2016fbr,MartinCamalich:2016wht,Ciuchini:2016weo,Capdevila:2017ert}. To
corroborate even more the cumbersome picture of the ``$P_{5}'$
anomaly'', two new independent measurements of this angular observable
(among others) have been recently released by ATLAS
\cite{ATLAS-CONF-2017-023} and CMS \cite{CMS-PAS-BPH-15-008}
collaborations, showing respectively an appreciable increase and
reduction of the tension between data and the SM prediction in
ref.~\cite{Descotes-Genon:2014uoa}, as reported by these experiments.

For the sake of completeness, one should also remark that other
smaller tensions have been around, concerning the measurement of
differential branching fractions of $B \to K \mu \mu\,$
\cite{Aaij:2014pli,Aaij:2016cb} and $B_{s} \to \phi \mu \mu$
\cite{Aaij:2015esa}. It is worth noting that, while for the latter
mode an explanation in terms of hadronic physics may be easily
conceivable, the theoretical computation of the former seems to be
under control \cite{Khodjamirian:2012rm}.

Quite surprisingly, a possible smoking gun for NP in rare $B$ meson
decays already came out in 2014, when the LHCb collaboration presented
for the first time the measurement of the ratio of branching fractions
\cite{Aaij:2014ora}:
\begin{eqnarray}
R_{{K}_{[1,6]}}  \ &\:\equiv\: &\ \frac{Br(B^{+} \to K^{+} \mu^{+} \mu^{-})}{Br(B^{+} \to K^{+} e^{+} e^{-})} \\ \nonumber
\ & = & \  0.745 ^{+ 0.090}_{-0.074}\pm 0.036  \ ,
\label{eq:RK}
\end{eqnarray}
where the subscript refers to the dilepton mass (denoted hereafter with $q^{2}$) range going from 1 to
6 GeV$^{2}$. This experimental value shows a deviation
of about $2.6\sigma$ with respect to the standard theoretical
prediction. Indeed, the SM value of $R_{K}$ in the bin provided by the
LHCb collaboration is expected to be equal to unity beyond the
percent level of accuracy \cite{Hiller:2003js,Bordone:2016gaq}. In fact, contrary
to observables such as $P_{5}'$, it is important to stress that
$R_{{K}}$ may be, in general, regarded as insensitive to QCD
effects  \cite{Hiller:2003js}. From the model building point of view, $R_K$ can certainly be
considered as quite informative, hinting at a UV completion of the SM
where Lepton Flavour Universality violation (LFUV) takes place in the
flavour-violating couplings of new heavy degrees of freedom,
e.g. leptoquarks and/or $Z'$ gauge bosons
\cite{Alonso:2014csa,Hiller:2014yaa,Ghosh:2014awa,Glashow:2014iga,Hiller:2014ula,Gripaios:2014tna,Sahoo:2015wya,Crivellin:2015lwa,Crivellin:2015era,Celis:2015ara,Alonso:2015sja,Greljo:2015mma,Calibbi:2015kma,Falkowski:2015zwa,Carmona:2015ena,Chiang:2016qov,Becirevic:2016zri,Feruglio:2016gvd,Megias:2016bde,Becirevic:2016oho,Arnan:2016cpy,Sahoo:2016pet,Alonso:2016onw,Hiller:2016kry,Galon:2016bka,Crivellin:2016ejn,GarciaGarcia:2016nvr,Cox:2016epl,Jager:2017gal,Megias:2017ove}. Most
importantly, the tantalizing correlation of this signature of LFUV
with the $P_{5}'$ anomaly, suggested by several global analyses
\cite{Beaujean:2013soa,Hurth:2013ssa,Altmannshofer:2014rta,Descotes-Genon:2015uva,Chobanova:2017ghn,Altmannshofer:2017fio}
has triggered different proposals of measurements of such effect in
the angular analysis of the $K^{*} \ell \ell$ channel
\cite{Capdevila:2016ivx,Serra:2016ivr}. Interestingly enough, an
analysis from the Belle collaboration aiming at separating the leptonic
flavours in $B \to K^{*} \ell \ell$~ \cite{Wehle:2016yoi}, shows a
consistent $\sim 2.6\sigma$ deviation from the SM prediction reported
in ref.~\cite{Descotes-Genon:2014uoa} in the dimuon leptonic final
state only. This is compatible with previous experimental findings
related only to the mode with muonic final states.

Sitting on similar theoretical grounds as $R_{K}$, another intriguing
ratio of $B$ decay branching fractions can be measured in the $K^{*}$
channel:
\begin{eqnarray}
R_{{K^{*}}_{[0.045,1.1]}} \ &\:\equiv\: &\ \frac{Br(B \to K^{*} \mu^{+} \mu^{-})}{Br(B \to K^{*} e^{+} e^{-})}  \\ \nonumber
\ & = & \ 0.660 ^{+ 0.110}_{-0.070}\pm 0.024 \ ,\\
R_{{K^{*}}_{[1.1,6]}} \ &=& \ 0.685 ^{+ 0.113}_{-0.069}\pm 0.047 \ .
\label{eq:RKstar}
\end{eqnarray}
These measurements for the low-$q^2$ bin and the central-$q^2$ one
have just been presented by the LHCb
collaboration~\cite{LHCb_RKstar}, pointing again to a discrepancy of
about $2\sigma$ with respect to the expected SM prediction -- again
equal to 1 to a very good accuracy for the central-$q^2$ bin and close to $0.9$ for the
low-$q^2$ one -- and yielding more than a $3\sigma$ deviation when naively combined
with the measurement of $R_K$.  Note that with higher degree of braveness
(or, depending on the taste of the reader, of unconsciousness),
the disagreement of the SM with precision $B$ physics may reach the exciting level of
$\gtrsim 5\sigma$ when one naively combines together the single
significances coming from $R_{K,K^{*}}$ ratios, $P_{5}'$ measurements
and the minor deviations observed in the other exclusive branching
fractions.

Given the excitement of these days for all the above hints of a
possible NP discovery in rare $B$ meson decays, in this work we take
our first steps towards a positive attitude in the search of a
definite BSM pattern aimed at addressing these $B$ anomalies. We
perform our study in a model-independent fashion, within the framework
of effective field theories for weak interactions
\cite{Buras:1992tc,Buras:1992zv,Ciuchini:1993vr}. In particular, in
section~\ref{sec:framework} we define the setup characterizing the
whole global analysis, presenting six different benchmark scenarios
for NP, together with a discussion about two different approaches in
the estimate of the hadronic uncertainties that can affect
quantitatively our final results. In section~\ref{sec:fit}, we list
all the experimental measurements we use to construct the likelihood
in our fit, and we discuss in detail our most important findings. The
latter are effectively depicted in
figures~\ref{fig:fig1}--\ref{fig:fig6}, and collected in
tables~\ref{tab:PMDpars}--\ref{tab:PDDobs} in ~\ref{sec:tab}. In section~\ref{sec:conclusions} we summarize
our conclusions.

\section{Theoretical Framework of the Analysis}
\label{sec:framework}

In this section we present the effective field theory framework at the
basis of this work and introduce the benchmark scenarios we focus on
for our study of NP effects in rare $B$ decays.  We then illustrate
the two distinct broad classes of assumptions that characterize our
global analysis: the case where we take an optimistic attitude towards
the estimate of hadronic uncertainty plaguing the amplitude of both
$B \to K^{*} \ell \ell / \gamma$ and $B_s \to \phi \ell \ell / \gamma$
channels, and a second one where we aim at providing a more
conservative approach.  All the results in section~\ref{sec:results}
will be classified under these two different setups.

\subsection{\textit{New Physics Benchmarks for $\Delta B = 1$}}
\label{sec:eft&NP}
Integrating out the heavy degrees of freedom, the resulting effective
Hamiltonian of weak interactions for $b \to s \gamma, s \ell \ell$
transitions involves the following set of dimension six operators
within the SM~\cite{Chetyrkin:1997gb}:
\begin{eqnarray}
 Q^p_1 &\:\:=\:\:& (\bar{s}_L\gamma_{\mu}T^a p_L)(\bar{p}_L\gamma^{\mu}T^ab_L)\,,\nonumber \\
 Q^p_2 &=& (\bar{s}_L\gamma_{\mu} p_L)(\bar{p}_L\gamma^{\mu}b_L)\,, \nonumber \\
 P_3 &=& (\bar{s}_L\gamma_{\mu}b_L)\sum\phantom{} _q(\bar{q}\gamma^{\mu}q)\,, \nonumber \\
 P_4 &=& (\bar{s}_L\gamma_{\mu}T^ab_L)\sum\phantom{}
         _q(\bar{q}\gamma^{\mu}T^aq) \,,\nonumber \\
 P_5 &=&
         (\bar{s}_L\gamma_{\mu1}\gamma_{\mu2}\gamma_{\mu3}b_L)\sum\phantom{}
         _q(\bar{q}\gamma^{\mu1}\gamma^{\mu2}\gamma^{\mu3}q)
         \,,\nonumber \\
 P_6 &=&
         (\bar{s}_L\gamma_{\mu1}\gamma_{\mu2}\gamma_{\mu3}T^ab_L)\sum\phantom{}
         _q(\bar{q}\gamma^{\mu1}\gamma^{\mu2}\gamma^{\mu3}T^aq)
         \,,\nonumber \\
 Q_{8g} &=& \frac{g_s}{16\pi^2}m_b\bar{s}_L\sigma_{\mu\nu}G^{\mu\nu}b_R \,,\nonumber \\
 Q_{7\gamma} &=&
                       \frac{e}{16\pi^2}m_b\bar{s}_L\sigma_{\mu\nu}F^{\mu\nu}b_R\,,
                       \nonumber \\
Q_{9V} &=&
          \frac{\alpha_{e}}{4\pi}(\bar{s}_L\gamma_{\mu}b_L)(\bar{\ell}\gamma^{\mu}\ell)\,,
          \nonumber \\
Q_{10A} &=&
           \frac{\alpha_{e}}{4\pi}(\bar{s}_L\gamma_{\mu}b_L)(\bar{\ell}\gamma^{\mu}\gamma^5\ell)  \,,
\label{eq:deltaB1}
\end{eqnarray}
where $\ell = e , \mu$, $p = u,c$ and we have neglected the chirally
suppressed SM dipoles.  The $\Delta B = 1$ effective Hamiltonian can
be casted in full generality as a combination of two distinct parts:
\begin{equation}
\mathcal{H}_\mathrm{eff}^{\Delta B = 1} =
\mathcal{H}_\mathrm{eff}^\mathrm{had} +
\mathcal{H}_\mathrm{eff}^\mathrm{sl+\gamma},
\label{eq:Heff}
\end{equation}
where, within the SM, the hadronic term involves the first seven
operators in eq.~(\ref{eq:deltaB1}):
\begin{eqnarray}
\mathcal{H}_\mathrm{eff}^\mathrm{had} \;\;&
= &\;\; \frac{4G_F}{\sqrt{2}} \Bigg[\sum_{p=u,c}\lambda_p\bigg(C_1 Q^{p}_1 +
C_2 Q^{p}_2\bigg) \\ \nonumber
\ & - & \ \lambda_t \bigg(\sum_{i=3}^{6} C_i P_i +
C_{8}Q_{8g} \bigg)\Bigg] \,,
\label{eq:H_had}
\end{eqnarray}
while the second piece includes the electromagnetic dipole and
semileptonic operators:
\begin{equation}
\mathcal{H}_\mathrm{eff}^\mathrm{sl+\gamma} = -
\frac{4G_F}{\sqrt{2}}\lambda_t\bigg( C_7Q_{7\gamma} + C_9Q_{9V} +
C_{10}Q_{10A} \bigg) \,,
\label{eq:H_sl}
\end{equation}
with $\lambda_{i}$ corresponding to the CKM combination
$V^{}_{ib} V^{*}_{is} $ for $i=u,c,t$ and where $C_{i=1,\dots,10}$ are
the Wilson coefficients (WCs) encoding the short-distance physics of
the theory. All the SM WCs in this work are evolved from the mass
scale of the W boson down to $\mu_{b}=4.8$ GeV, using
state-of-the-art perturbative QCD and QED calculations for the
matching conditions \cite{Bobeth:1999mk,Gambino:2001au,Misiak:2004ew}
and the anomalous dimension matrices
\cite{Bobeth:2003at,Gambino:2003zm,Misiak:2004ew,Huber:2005ig}
relevant for the processes considered in this analysis.

While a general UV completion of the SM may enter in the effective
couplings present in both pieces of eq.~(\ref{eq:Heff}), general
NP effects in $b \to s \gamma, s \ell \ell$ can be
phenomenologically parametrized as shifts of the Wilson coefficients
of the electromagnetic and semileptonic operators at the typical scale
of the processes, $\mu_{b}$. In particular, the most general basis for
NP effects in radiative and (semi)leptonic $B$ decays can be enlarged
by the presence of scalar, pseudo-scalar and tensorial semileptonic
operators, together with right-handed quark currents as the analogue
of $Q_{7\gamma} ,Q_{9V},Q_{10A}$ SM operators
\cite{Jager:2012uw,Aebischer:2015fzz}.  In this work, motivated by
previous interesting findings concerning LFUV
\cite{Altmannshofer:2014rta,Descotes-Genon:2015uva,Chobanova:2017ghn}
and the measurement of $R_{K}$ and $R_{K^{*}}$, we focus on the
contributions of NP appearing as shifts of the SM WCs related to the
electromagnetic dipole and semileptonic operators with left-handed
quark currents only. A comprehensive analysis with different chiral
structures as well as a more general effective theory framework will
be presented elsewhere~\cite{future}. Furthermore, we restrict
ourselves to CP-conserving effects, taking NP WCs to be real.

For NP in semileptonic operators we discriminate between couplings to muon
and electron fields both in the axial and vector leptonic currents.
We characterize our phenomenological analysis for NP through six
different benchmark scenarios, studying the impact of combinations of
the following NP WCs :
\begin{flushleft}
\begin{itemize}
\item[\textbf{(I)}] $C^{NP}_{9,\mu}$ and $C^{NP}_{9,e}$ varied in the range $[-4,4]$,
  i.e. adding to the SM two NP parameters;
\item[\textbf{(II)}] $C^{NP}_{9,\mu}$ and $C^{NP}_{10,\mu} $ varied in
  the range $[-4,4]$, adding to the SM again two NP parameters;
\item[\textbf{(III)}] $C^{NP}_{9,\mu}$ and $C^{NP}_{9,e}$ varied in the range $[-4,4]$,
and $C^{NP}_{7}$ varied in the range $[-0.5,0.5]$,
i.e. a scenario with three NP parameters;
\item[\textbf{(IV)}] $C^{NP}_{10,\mu}$ and $C^{NP}_{10,e}$ varied in the range $[-4,4]$,
and $C^{NP}_{7}$ varied in the range $[-0.5,0.5]$,
 i.e. adding again to the SM three NP
  parameters;
\item[\textbf{(V)}] $C^{NP}_{9,\mu} = - C^{NP}_{10,\mu}$ and $C^{NP}_{9,e} = - C^{NP}_{10,e}$
varied in the range $[-4,4]$,
and $C^{NP}_{7}$ varied in the range $[-0.5,0.5]$, i.e. a
NP scenario again described by three different parameters.
\item[\textbf{(VI)}] $C^{NP}_{7}$, $C^{NP}_{9,\mu}$, $C^{NP}_{9,e}$,
  $C^{NP}_{10,\mu}$ and $C^{NP}_{10,e}$ varied simultaneously in the respective ranges defined
  above, i.e. a NP scenario described by five different
  parameters.
\end{itemize}
\end{flushleft}

We remark that while benchmarks \textbf{(I)} and \textbf{(II)} have
been already studied in literature, none of the other cases has been
analyzed so far. In particular, NP scenarios \textbf{(III)} and
\textbf{(IV)} allow us to study, for the first time, the interesting
impact of a NP radiative dipole operator in combination with
vector-like and axial-like LFUV effects generated by NP. Most
interestingly, scenario \textbf{(V)} allows us to explore the
correlation $C_{9}^{NP}=-C_{10}^{NP}$, possibly hinting at a
$SU(2)_{L}$ preserving BSM theory. As an additional interesting case
to explore, we eventually generalize to simultaneously nonvanishing
$C^{NP}_{7}$, $C^{NP}_{9,\mu}$, $C^{NP}_{9,e}$,
  $C^{NP}_{10,\mu}$ and $C^{NP}_{10,e}$ in case \textbf{(VI)}.

We wish to stress that all of the six benchmarks defined above will
be studied for the first time under two different approaches in the
estimate of QCD hadronic power corrections, as presented in
next section.

\subsection{\textit{Treatment of the Hadronic Uncertainties}}
\label{sec:hadronic}

In our previous works~\cite{Ciuchini:2015qxb,Ciuchini:2016weo,Ciuchini:2017gva}, we
went into considerable detail on the treatment of hadronic
contributions in the angular analysis of $B\to K^*\ell\ell$. Our
approach there was to study how large these contributions can be
assuming that the LHCb data on branching fractions and angular
distributions of these decay modes could be described within the
SM. For that purpose we considered four scenarios for the hadronic
contributions, with increasing theoretical input from the
phenomenological analysis presented in
ref.~\cite{Khodjamirian:2010vf}.
The underlying functional form that
we used for the hadronic contribution was given by:
\begin{eqnarray}
 h_\lambda(q^2)  &\;\;=\;\;& \frac{\epsilon^*_\mu(\lambda)}{m_B^2} \int d^4x e^{iqx} \langle \bar K^* \vert T\{j^{\mu}_\mathrm{em} (x)
 \mathcal{H}_\mathrm{eff}^\mathrm{had} (0)\} \vert \bar B \rangle \nonumber\\
 &\;\;=\;\;& h_\lambda^{(0)} + \frac{q^2}{1\,\mathrm{GeV}^2}
         h_\lambda^{(1)} + \frac{q^4}{1\, \mathrm{GeV}^4} h_\lambda^{(2)} \,,
 \label{eq:hlambda}
\end{eqnarray}
where we fitted for the complex, helicity dependent, coefficients
$h^{(i)}_\lambda$, $(i=0,1,2)$ and $(\lambda=0,+,-)$ using the data
and the phenomenological model in \cite{Khodjamirian:2010vf}. Since
$h_0$ enters the decay amplitude with an additional factor of
$\sqrt{q^2}$ with respect to $h_\pm$, we drop $h_0^{(2)}$ in our
analysis.

In this work we proceed to study the possible existence of NP
contributions in semileptonic and radiative $b\to s$ decays which
requires a re-evaluation of the hadronic uncertainties. For the sake
of simplicity, to address hadronic contributions we use the same
functional parameterization as given in
eq.~(\ref{eq:hlambda}). However, we limit ourselves to only two
hadronic models. The first, corresponding to the most widely used
assumption, relies completely on the phenomenological model
in~\cite{Khodjamirian:2010vf} below $q^2 < 4m_c^2$. The second is a
more conservative approach, where we impose the latter only in the
large recoil region at $q^2\le1$ GeV$^2$ while letting the data drive
the hadronic contributions in the higher invariant mass region. We
will refer to the first approach as phenomenological model driven
(PMD) and the second as phenomenological and data driven (PDD). In our
fit we vary the $h^{i}_\lambda$ parameters over generous ranges.  More
detailed discussion on these can be found
in~\cite{Ciuchini:2015qxb,Ciuchini:2016weo}.

In the present analysis we also need to address modes that were not
considered in our previous works, namely $B\to K\ell\ell$,
$B_{s}\to \phi\ell\ell$ and $B_{s}\to \phi\gamma$. The decay
$B\to K\ell\ell$ has been studied in detail
in~\cite{Khodjamirian:2012rm}, where the authors show that the
hadronic uncertainties are smaller than in $B\to K^*\ell\ell$. A
comparison of the LCSR estimate of the soft gluon contribution and the
QCDF estimate of the hard gluon contribution reveals that the soft
gluon exchange is subdominant with respect to QCDF hard gluon
exchange. Therefore, although in principle the same concerns on the
soft gluon contribution we raised for $B \to K^*$ apply also in this
case, in practice the overall effect of soft gluons can be reasonably
neglected. In our computation we therefore only include hard gluon
exchange computed using the QCDF formalism in
ref.~\cite{Beneke:2001at}.

The long distance contributions for $B_s\to \phi\ell\ell$ and
$B_s\to \phi\gamma$ follow a similar theoretical derivation as those
for $B\to K^*\ell\ell$ and $B\to K^*\gamma$, respectively, barring the
fact that the spectator quark in the former is different from that in
the latter. No theoretical estimates of power corrections to the
infinite mass limit are available for the
$B_s \to \phi \ell \ell / \gamma$ decays and one has to rely on the
ones for the $B\to K^* \ell \ell / \gamma$ decays to get a handle on
the long distance contributions. The spectator quark effects can come
through the hard spectator scattering involving matrix elements of
$Q_{2}$, $P_6$ and $Q_{8g}$ computable in QCD
factorization~\cite{Beneke:2001at} which we include in our
computation. However, we do not include the sub-leading, and
numerically small, QCDF power corrections to spectator scattering
involving $Q_{8g}$~\cite{Kagan:2001zk,Feldmann:2002iw,Beneke:2004dp}
and contributions to weak spectator scattering involving $Q_{8g}$
beyond QCDF computed in
LCSR~\cite{Ball:2006eu,Dimou:2012un,Lyon:2013gba}.  The effect of the
difference in all these spectator contributions is expected to be low
firstly because they are numerically small and, secondly, because the
effect is proportional to the small flavour $SU(3)$
breaking. Different approaches in relating the long distance
contributions in the $B\to K^* \ell \ell / \gamma$ channels to the
ones in the $B \to \phi \ell \ell / \gamma$ channels have been used in
the
literature~\cite{Altmannshofer:2014rta,Paul:2016urs,Descotes-Genon:2015uva},
which vary in the degree of correlation between the two. While
Ref~\cite{Descotes-Genon:2015uva} uses uncorrelated hadronic
uncertainties, refs.~\cite{Altmannshofer:2014rta,Paul:2016urs} have
left the two contributions highly correlated noting that the spectator
contribution is expected to be numerically small. We take an approach
similar to the the latter considering the insensitivity of the current
data to such effects and use the same value of power corrections in
$B \to K^*$ and $B_s \to \phi$ amplitudes, even though this choice
pertains to a quite oversimplifying optimistic attitude.  We leave a
more detailed analysis of this assumption by relaxing the correlation
between the hadronic contributions in the two modes to a future
work~\cite{future}.

\section{Bayesian Fit of the Dipole and Semileptonic Operators}
\label{sec:fit}

\subsection{\textit{Experimental Information Considered}}
In this section we discuss the experimental measurements we use in our
fit. Please note that for the exclusive modes we make use of
measurements in the large recoil region only. Our choice harbours on
the fact that the QCD long distance effects in the low recoil region
are substantially different from the large recoil
regime~\cite{Grinstein:2004vb,Bobeth:2010wg,Beylich:2011aq,Bobeth:2011gi}
and would require a dedicated analysis. For the fit in this study we
consider the following experimental information:

\begin{itemize}
\item $B \to K^* \ell \ell$\\
  For the $B \to K^* \mu \mu$ channel we use the LHCb measurements of
  CP-averaged angular observables extracted by means of the unbinned
  maximum likelihood fit, along with the provided correlation
  matrix~\cite{Aaij:2015oid}. Moreover, we employ the recent results
  for CP-averaged angular observables from
  ATLAS~\cite{ATLAS-CONF-2017-023} and the ones measured by
  CMS~\cite{Khachatryan:2015isa,CMS-PAS-BPH-15-008}\footnote{For all
    CMS data we use the 7, 8 TeV combined results, which can be found
    in
    \url{https://twiki.cern.ch/twiki/bin/view/CMSPublic/PhysicsResultsBPH13010}
   .}
  as well. Finally, we use the CP-averaged optimized angular
  observables recently measured by
  Belle~\cite{Wehle:2016yoi}\footnote{Belle measures the
    $B^0 \to K^{*0} \mu \mu$ and $B^+ \to K^{*+} \mu \mu$
    channels together, without providing the mixture ratio. On the
    theoretical side, we can therefore use these measurements under
    the approximation that QCD power corrections differentiating the
    amplitudes of the two channels are small. We have numerically
    checked that the impact of known QCD power
    corrections~\cite{Beneke:2001at} is indeed at the percent level in
    the observables of interest.}. Regarding the differential
  branching fractions, we use the recently updated measurements from
  LHCb~\cite{Aaij:2016flj} and the ones from
  CMS~\cite{Khachatryan:2015isa}. For the $B \to K^* e e$ channel we
  consider the LHCb results from~\cite{Aaij:2015dea} and the Belle
  results from~\cite{Wehle:2016yoi}. $R_{K^*}$ observable is considered according
  to the recently presented measurements by LHCb~\cite{LHCb_RKstar} in both the low-$q^{2}$
  and central-$q^{2}$ bins, see also eq.~(\ref{eq:RKstar}).

  Our theoretical predictions are computed in the helicity basis,
  whose relevant expressions can be found in~\cite{Jager:2012uw}; the
  same framework is employed to study $B \to K^* \gamma$,
  $B_s \to \phi \mu \mu$, $B_s \to \phi \gamma$ and
  $B \to K \ell \ell$ channels. For the latter, we use the full set of
  form factors extrapolated from the lattice results, along with the
  provided correlation matrix~\cite{Bailey:2015dka}; for the remaining
  channels, we use the full set of form factors estimated combining
  LCSR and lattice results, along with the correlation
  matrices~\cite{Straub:2015ica}. For the factorizable and
  non-factorizable QCD power corrections, we refer to
  Sec.~\ref{sec:hadronic}.
\item $B \to K^* \gamma$\\
  We include in our analysis the HFAG average for the branching
  fractions from~\cite{Amhis:2016xyh}.
\item $B_s \to \phi \mu \mu$\\
  We consider the LHCb CP-averaged angular observables and
  differential branching fractions measurements, along with the
  provided correlation matrix~\cite{Aaij:2015esa}.
\item $B_s \to \phi \gamma$\\
  We use the LHCb measurement of the branching fraction
  from~\cite{Aaij:2012ita}.
\item $B \to K \ell \ell$\\
  We employ the LHCb measurement of $B \to K e e$ differential
  branching fraction and $R_K$ from~\cite{Aaij:2014ora}.
\item $B \to X_s \gamma$\\
  We use the HFAG average from~\cite{Amhis:2016xyh}. We perform our
  theoretical computation at NNLO in $\alpha_s$ and NLO in
  $\alpha_{em}$, following ref.~\cite{Misiak:2015xwa} and references
  therein.
\item $B_s \to \mu \mu$\\
  We consider the latest measurement from LHCb~\cite{Aaij:2017vad} and
  do not consider the measurement from CMS~\cite{Chatrchyan:2013bka},
  which has the same central value of LHCb, but larger
  uncertainty. Moreover, we chose not to use results for
  $B_d \to \mu \mu$, since there are only upper bounds for this decay
  channel so far~\cite{Chatrchyan:2013bka,Aaij:2017vad}.  Our
  theoretical predictions include NLO EW corrections, as well as NNLO
  QCD correction, following the detailed expressions obtained
  in ref.~\cite{Bobeth:2013uxa}.
\end{itemize}

\subsection{\textit{Results of the Global Fit}}
\label{sec:results}

In this section we present the main results of our work. We perform
this study using \HEPfit~\cite{HEPfit} relying on its Markov Chain
Monte Carlo based Bayesian analysis framework implemented with
BAT~\cite{Caldwell:2008fw}. We fit to the data using 16 real free
parameters that characterize the non-factorizable power corrections,
as was done in~\cite{Ciuchini:2015qxb}, along with the necessary set
of NP WCs.  We assign to the hadronic parameters and the NP WCs flatly
distributed priors in the relevant ranges mentioned in
section~\ref{sec:framework}. The remaining parameters used in the fit
are listed in table~\ref{Tab:SM}. To better compare different
scenarios, we use the Information Criterion \cite{IC,MR2027492},
defined as
\begin{equation}
  \label{eq:IC}
  \mathit{IC} = -2 \overline{\log L} + 4 \sigma^2_{\log L}\,,
\end{equation}
where $\overline{\log L}$ is the average of the log-likelihood and
$\sigma^2_{\log L}$ is its variance. The second term in eq.~(\ref{eq:IC}) takes into account
the effective number of parameters in the model, allowing for a
meaningful comparison of models with different number of parameters.
Preferred models are expected to
give smaller $IC$ values.

\begin{table*}[t!]
\centering
\begin{tabular}{|c|c|c|c|}
\hline
\textbf{Parameters} & \textbf{Mean Value} & \textbf{Uncertainty} & \textbf{Reference} \\[1mm]
\hline
$\alpha_{s}(M_{Z})$ & $0.1181$ & $ 0.0009 $ & \cite{Agashe:2014kda,deBlas:2016ojx}\\
$\mu_{W}$ (GeV) & $80.385$ &  $-$ & \\
$m_{t}$ (GeV) & $173.34$ & $0.76$ & \cite{ATLAS:2014wva}\\
$m_{c}(m_c)$ (GeV) & $1.28$ & $0.02$ & \cite{Lubicz}\\
$m_{b}(m_b)$ (GeV) & $4.17$ &  $0.05$ & \cite{Sanfilippo:2015era}\\
$f_{B_{s}}$ (MeV) & $226$ & $5$ & \cite{Aoki:2013ldr}\\
$f_{B_{s}}/ f_{B_{d}}$ & $1.204$ & $0.016$ & \cite{Aoki:2013ldr}\\
$\Delta \Gamma_s/\Gamma_s$ & $0.129$ & $0.009$ & \cite{Amhis:2016xyh} \\
$\lambda $ & $0.2250$ & $0.0006$ & \cite{Bona:2006ah,UTfit}\\
$A $ & $0.829$ & $0.012$ & \cite{Bona:2006ah,UTfit}\\
$\bar{\rho} $ & $0.132$ & $0.018$ & \cite{Bona:2006ah,UTfit}\\
$\bar{\eta} $ & $0.348$ & $0.012$ & \cite{Bona:2006ah,UTfit}\\
$f_{K^*,\vert\vert}$ (MeV) & $204$ & $7$ & \cite{Straub:2015ica}\\
$f_{K^*,\perp}(1\mathrm{GeV})$ (MeV) & $159$ & $6$ &\cite{Straub:2015ica} \\
$f_{\phi,\vert\vert}$ (MeV) & $233$ & $4$ &\cite{Straub:2015ica} \\
$f_{\phi,\perp}(1\mathrm{GeV})$ (MeV) & $191$ & $4$ & \cite{Straub:2015ica}\\[0.6mm]
\hline
&&&\\[-3mm]
$\lambda_{B}$ (MeV) & $350$  & $ 150$ & \cite{Bosch:2001gv}\\
$a_{1}(\bar{K}^{*})_{\perp, \, ||}$ & $0.04$ &  $0.03$ & \cite{Ball:2005vx}\\
$a_{2}(\bar{K}^{*})_{\perp, \, ||}$ & $0.05$ &  $ 0.1$ & \cite{Ball:2006nr}\\
$a_{2}(\phi)_{\perp, \, ||}$ & $0.23$ &  $ 0.08$ & \cite{Ball:2007rt}\\
$a_{1}(K)$ & $0.06$ &  $0.03$ & \cite{Ball:2005vx}\\
$a_{2}(K)$ & $0.115$ &  $ -$ & \cite{Ball:2004ye}\\
\hline
\end{tabular}
\caption{\it{Parameters used in the analysis. The Gegenbauer
    parameters and $\lambda_B$ have flat priors with half width
    reported in the third column. The remaining ones have Gaussian
    priors. Meson masses, lepton masses, $s$-quark mass and
    electroweak couplings are fixed at the PDG value
    \cite{Agashe:2014kda}.}}
 \label{Tab:SM}
\end{table*}

\begin{figure}[!t]
\centering
\subfigure{\includegraphics[width=.45\textwidth]{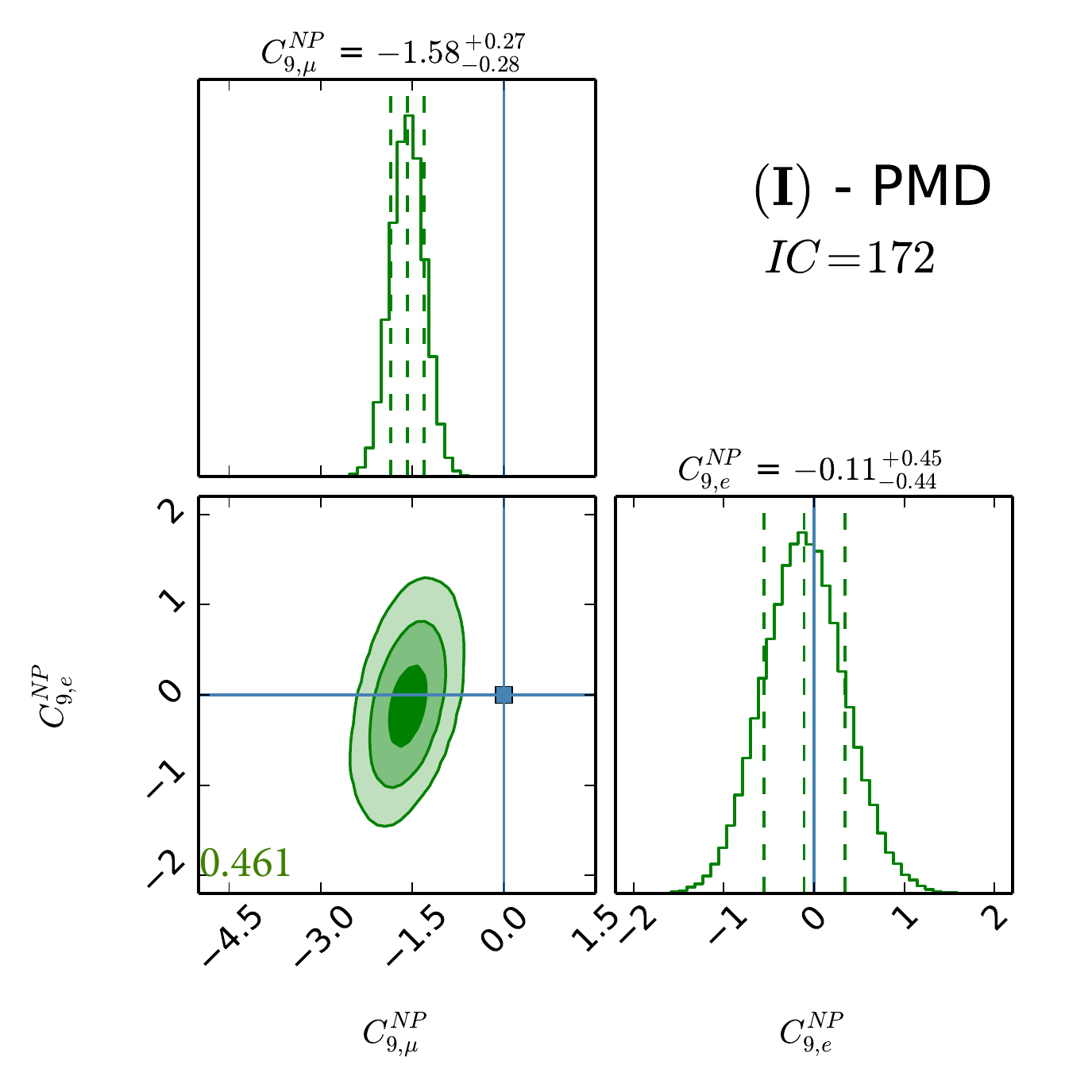}}
\subfigure{\includegraphics[width=.45\textwidth]{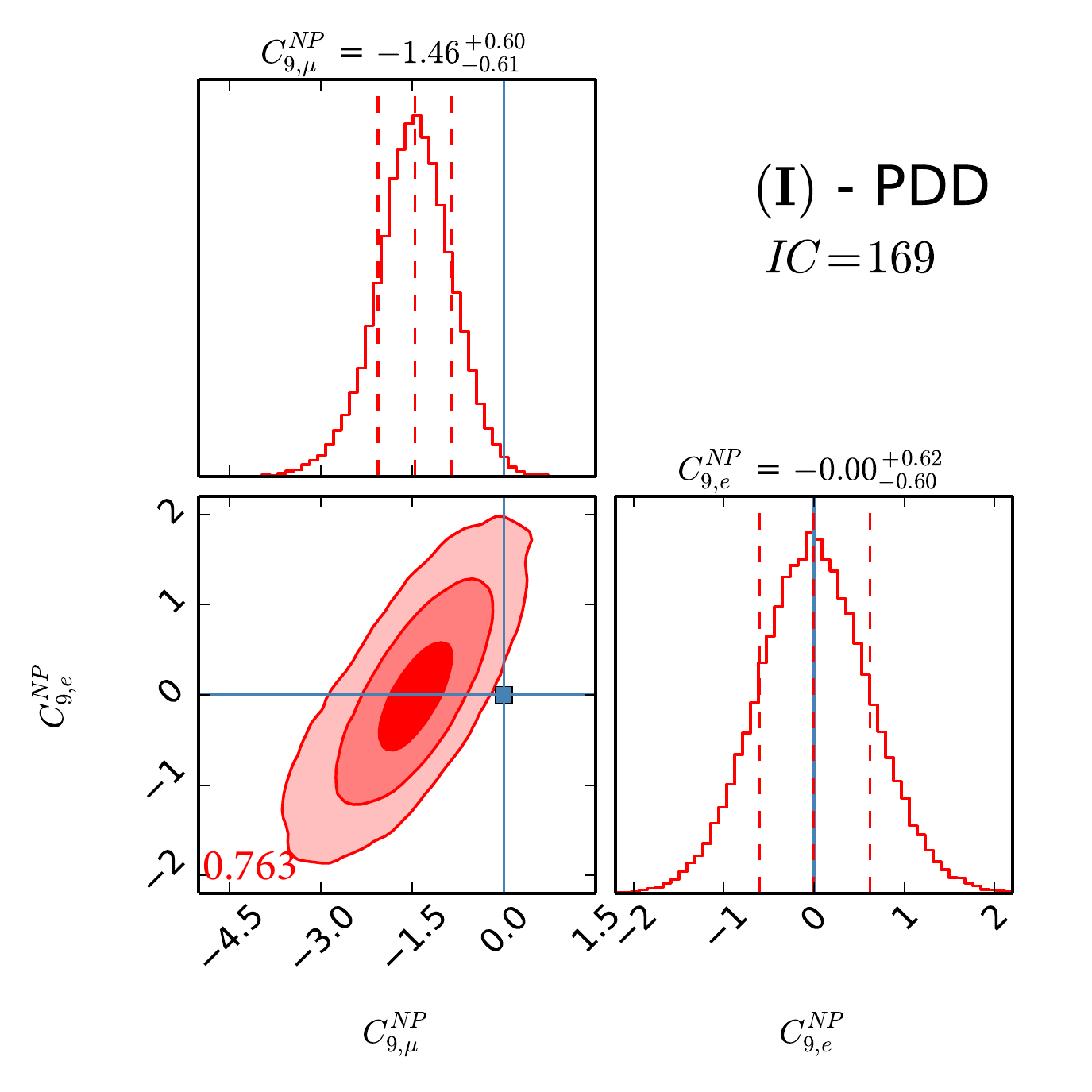}}
\caption{\textit{The two NP parameter fit using $C^{NP}_{9,\mu}$ and
    $C^{NP}_{9,e}$. Here and in the following, the left green panel
    shows the results for the PMD approach and the right red panel
    shows that for the PDD one. In the 1D distributions we show the
    $16^{th}$, $50^{th}$ and $84^{th}$ percentile marked with the
    dashed lines. In the correlation plots we show the 1, 2 and
    $3\sigma$ contours in decreasing degrees of transparency. The blue
    square and lines identify the values of the NP WCs in the SM
    limit. The numbers at the bottom left corner of the 2D plots refer
    to the correlation. We also report $IC$ values for the two
    approaches (see eq.~\ref{eq:IC}). Preferred models are expected
    to give smaller $IC$ values.}}
\label{fig:fig1}
\end{figure}

\begin{figure}[!t!]
\centering
\subfigure{\includegraphics[width=.45\textwidth]{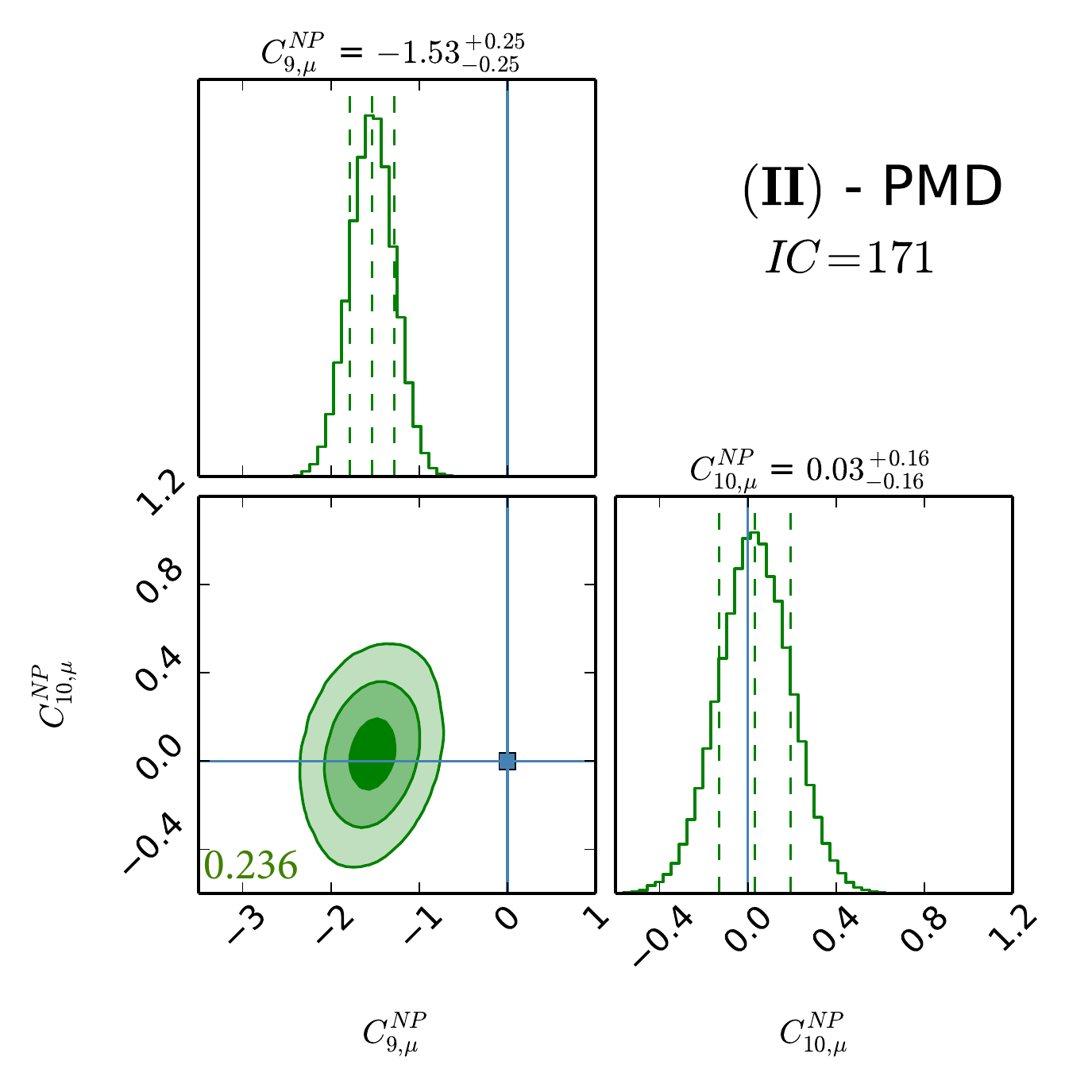}}
\subfigure{\includegraphics[width=.45\textwidth]{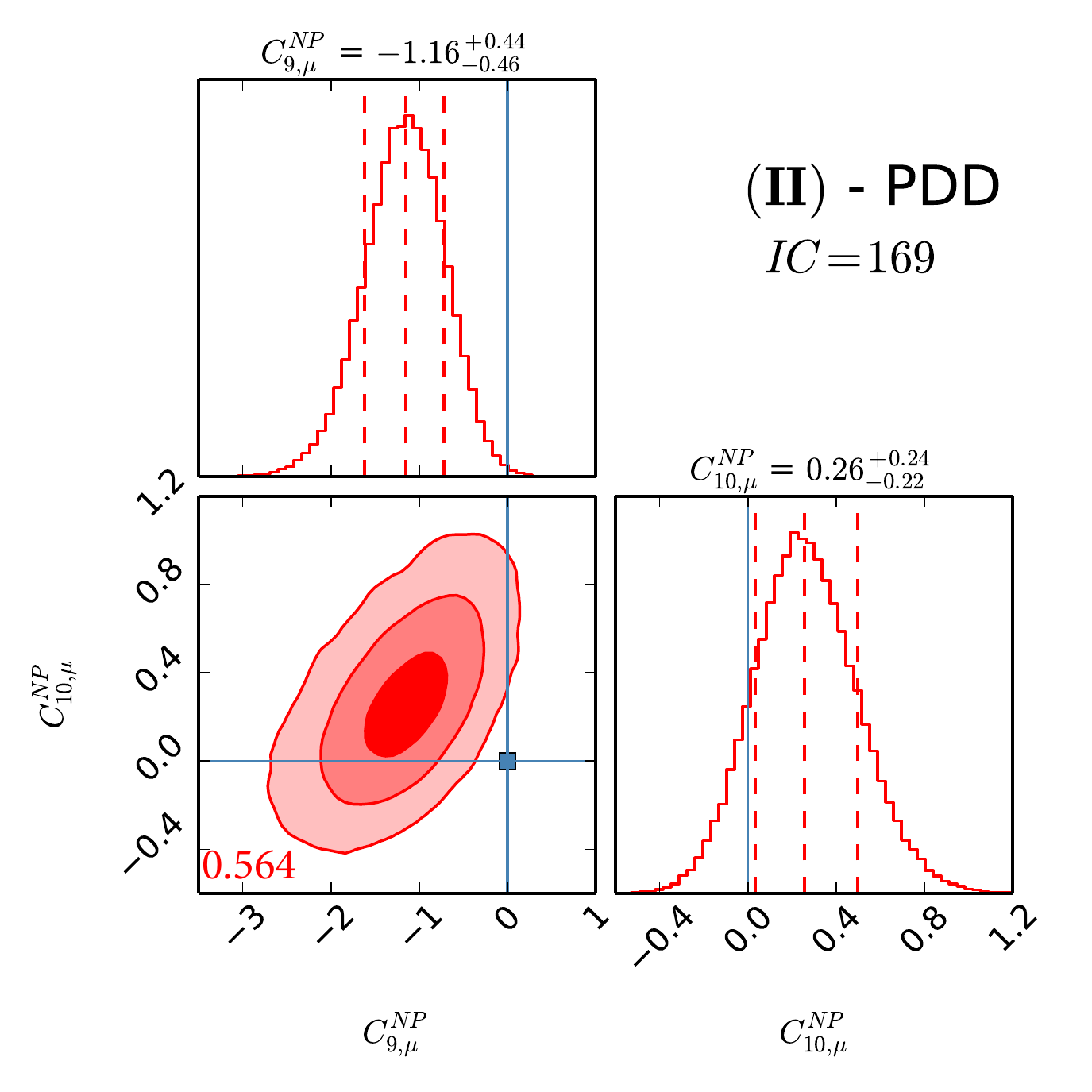}}
\caption{\textit{The two NP parameter fit using $C^{NP}_{9,\mu}$ and
    $C^{NP}_{10,\mu}$. See caption of figure \ref{fig:fig1} for the colour coding and further details.}}
\label{fig:fig2}
\end{figure}

\begin{figure}[!t!]
\centering
\subfigure{\includegraphics[width=.45\textwidth]{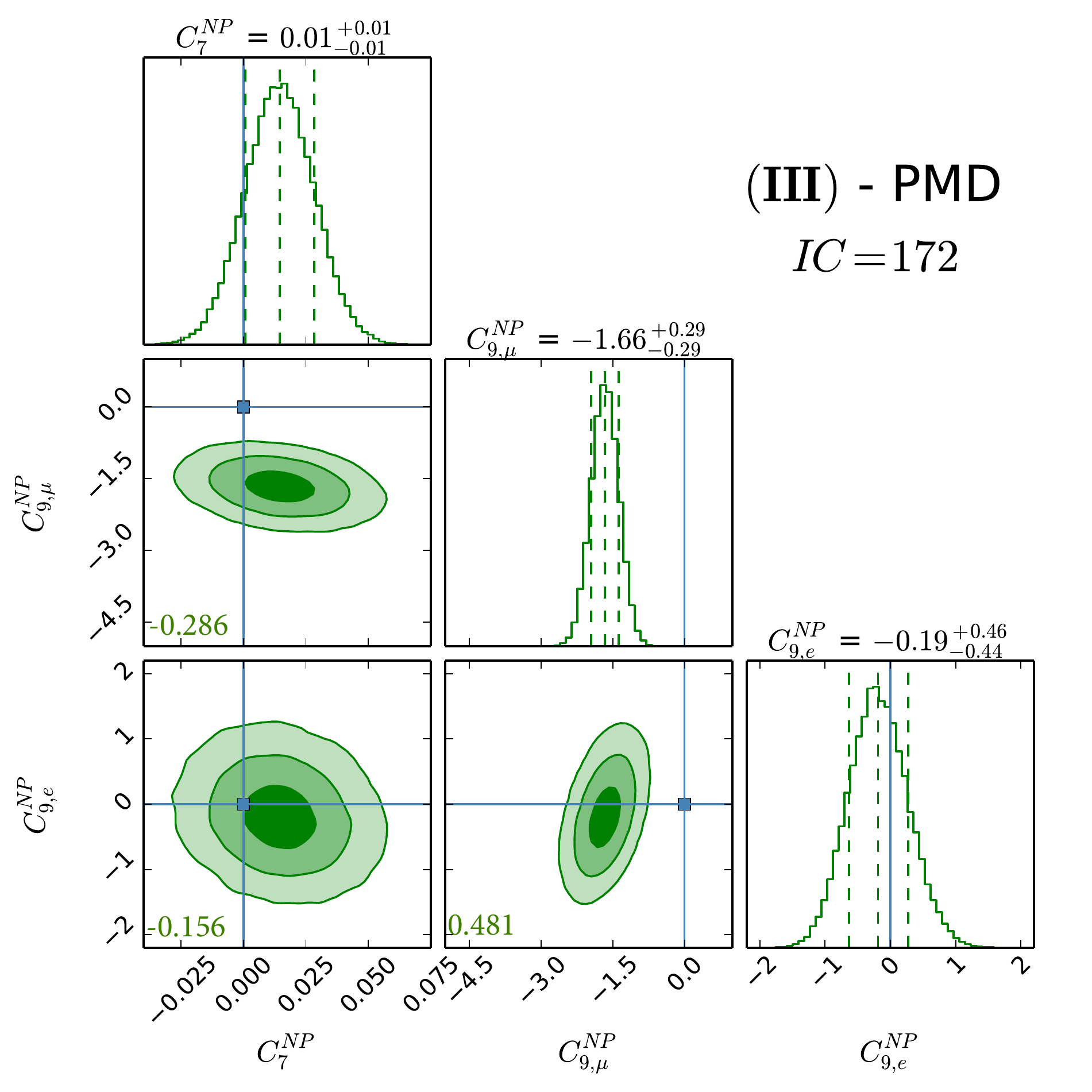}}
\subfigure{\includegraphics[width=.45\textwidth]{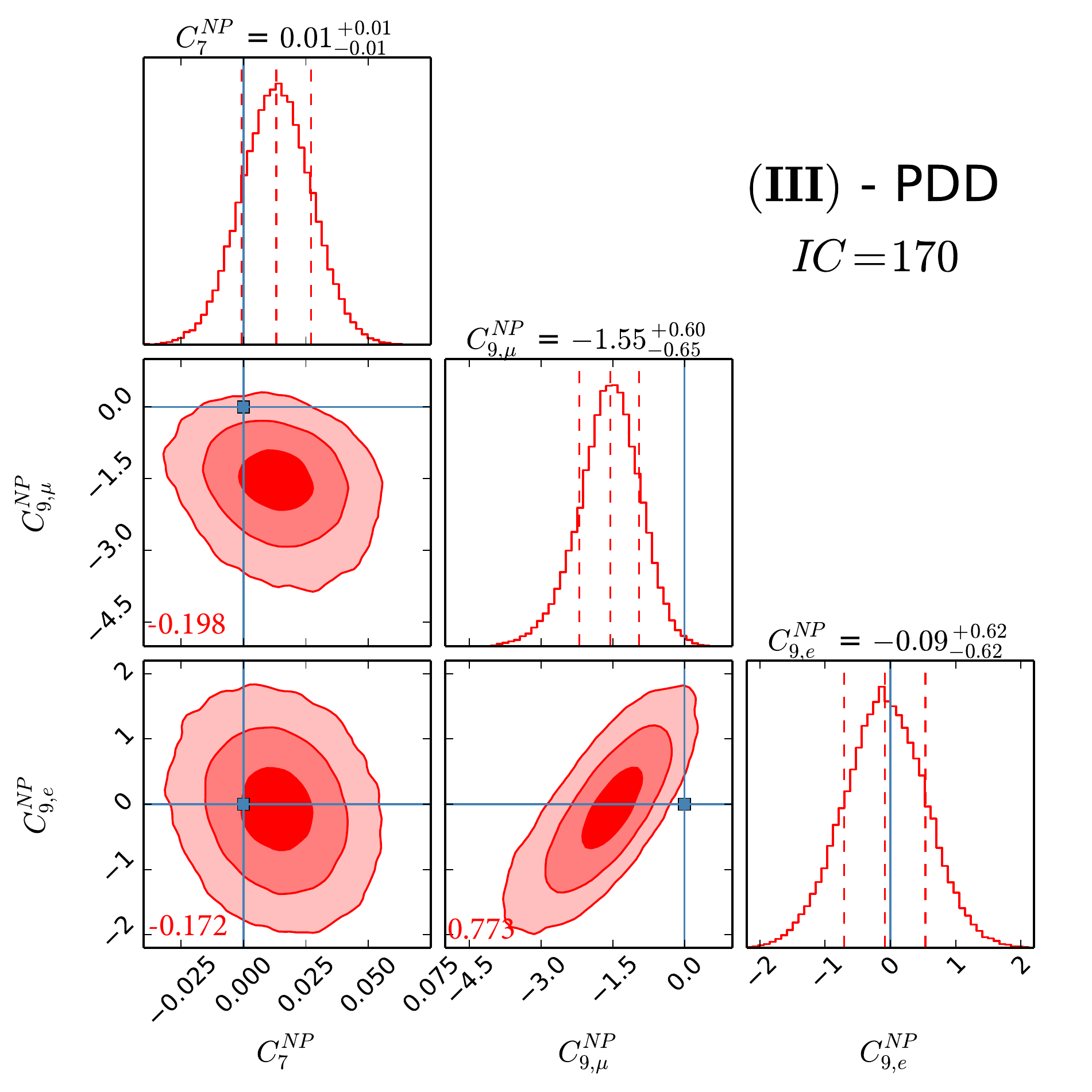}}
\caption{\textit{The three NP parameter fit using $C^{NP}_{7}$,
    $C^{NP}_{9,\mu}$ and $C^{NP}_{9,e}$. See caption of figure
    \ref{fig:fig1} for the colour coding and further details.}}
\label{fig:fig3}
\end{figure}

\begin{figure}[!ht!]
\centering
\subfigure{\includegraphics[width=.45\textwidth]{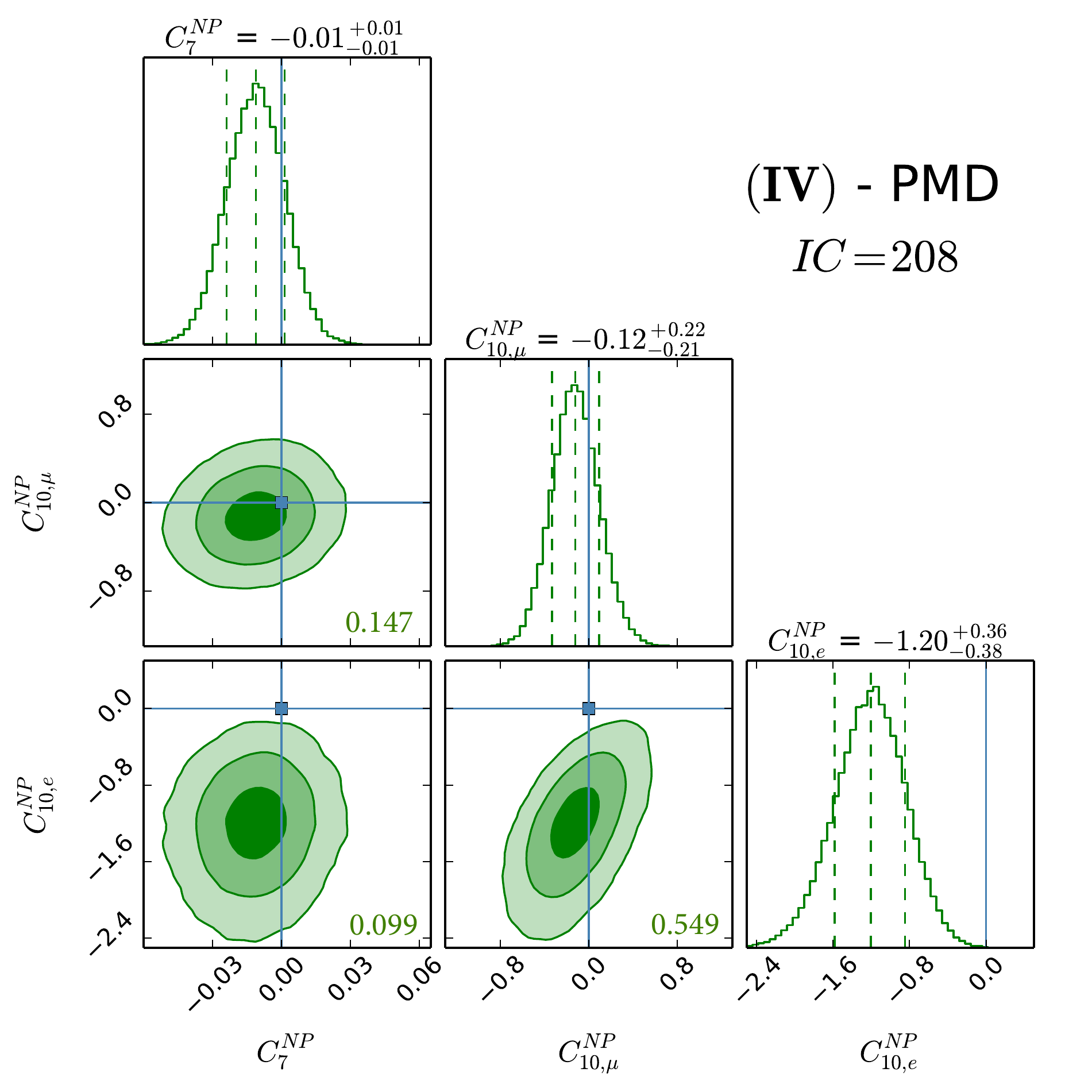}}
\subfigure{\includegraphics[width=.45\textwidth]{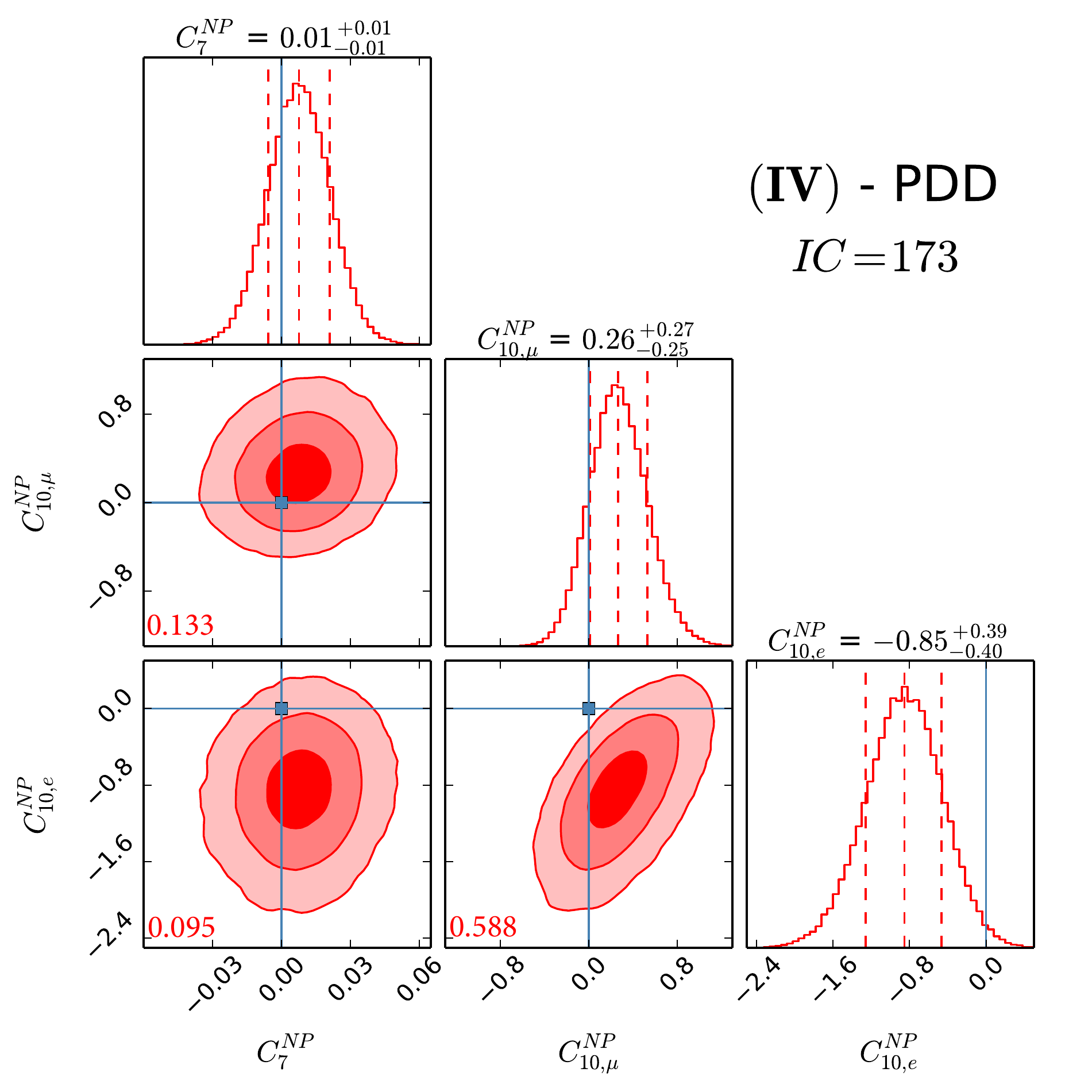}}
\caption{\textit{The three NP parameter fit using $C^{NP}_{7}$,
    $C^{NP}_{10,\mu}$ and $C^{NP}_{10,e}$. See caption of figure
    \ref{fig:fig1} for the colour coding and further details.}}
\label{fig:fig4}
\end{figure}

\begin{figure}[!ht!]
\centering
\subfigure{\includegraphics[width=.45\textwidth]{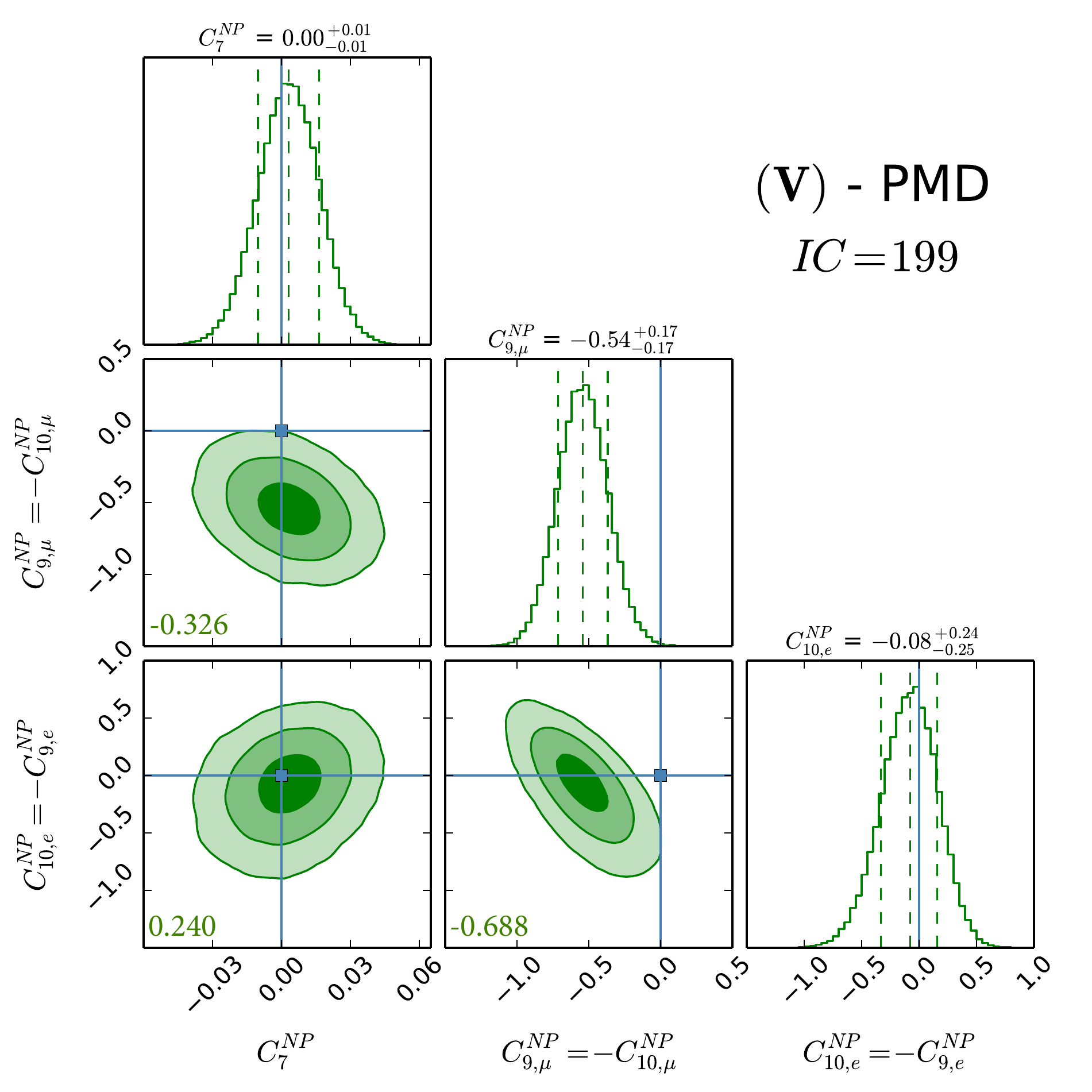}}
\subfigure{\includegraphics[width=.45\textwidth]{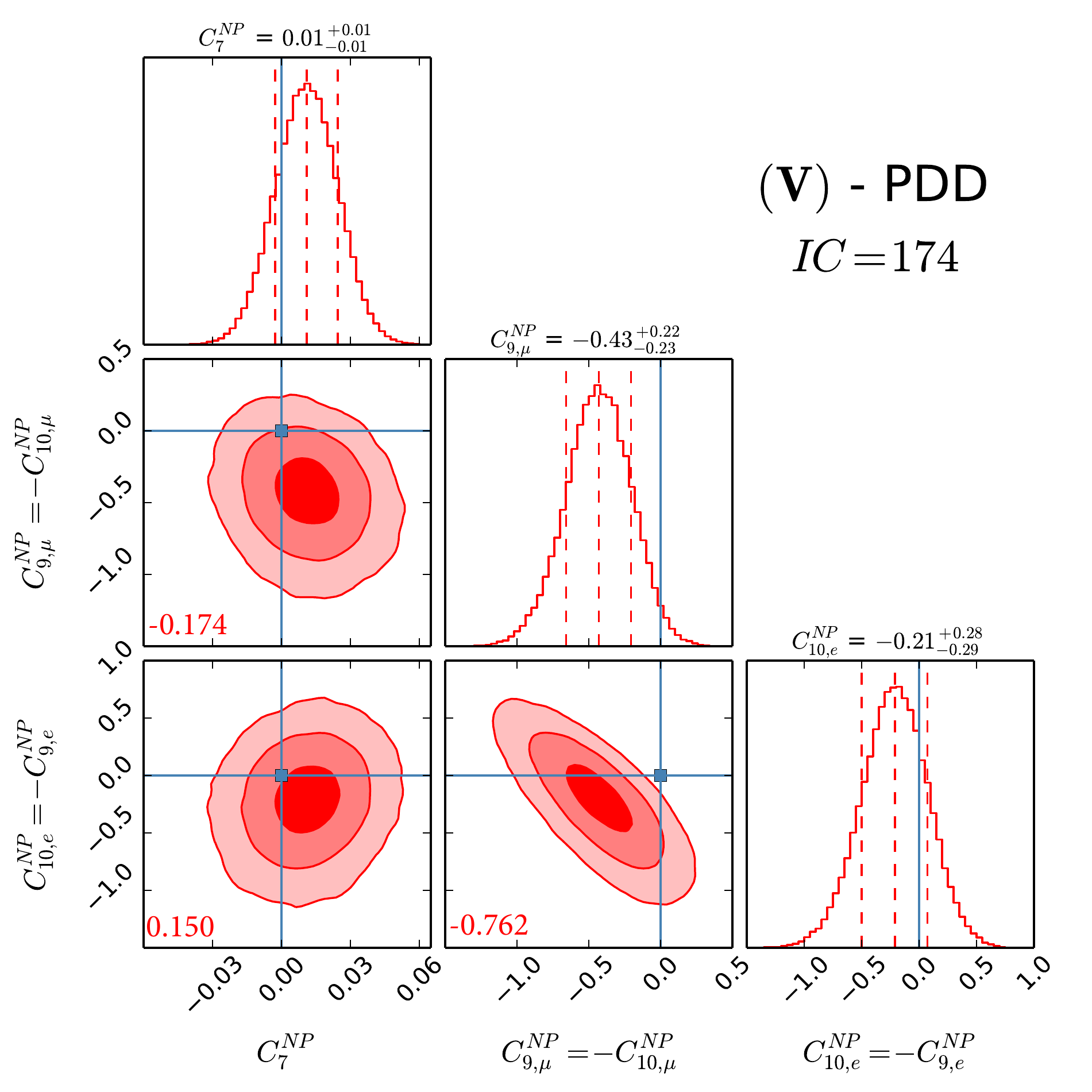}}
\caption{\textit{The three NP parameter fit using $C^{NP}_{7}$,
    $C^{NP}_{9,\mu}$, $C^{NP}_{9,e}$ and
    $C^{NP}_{10,\mu,e}=-C^{NP}_{9,\mu,e}$. See caption of figure
    \ref{fig:fig1} for the colour coding and further details.}}
\label{fig:fig5}
\end{figure}

\begin{figure}[!ht!]
\centering
\subfigure{\includegraphics[width=.5\textwidth]{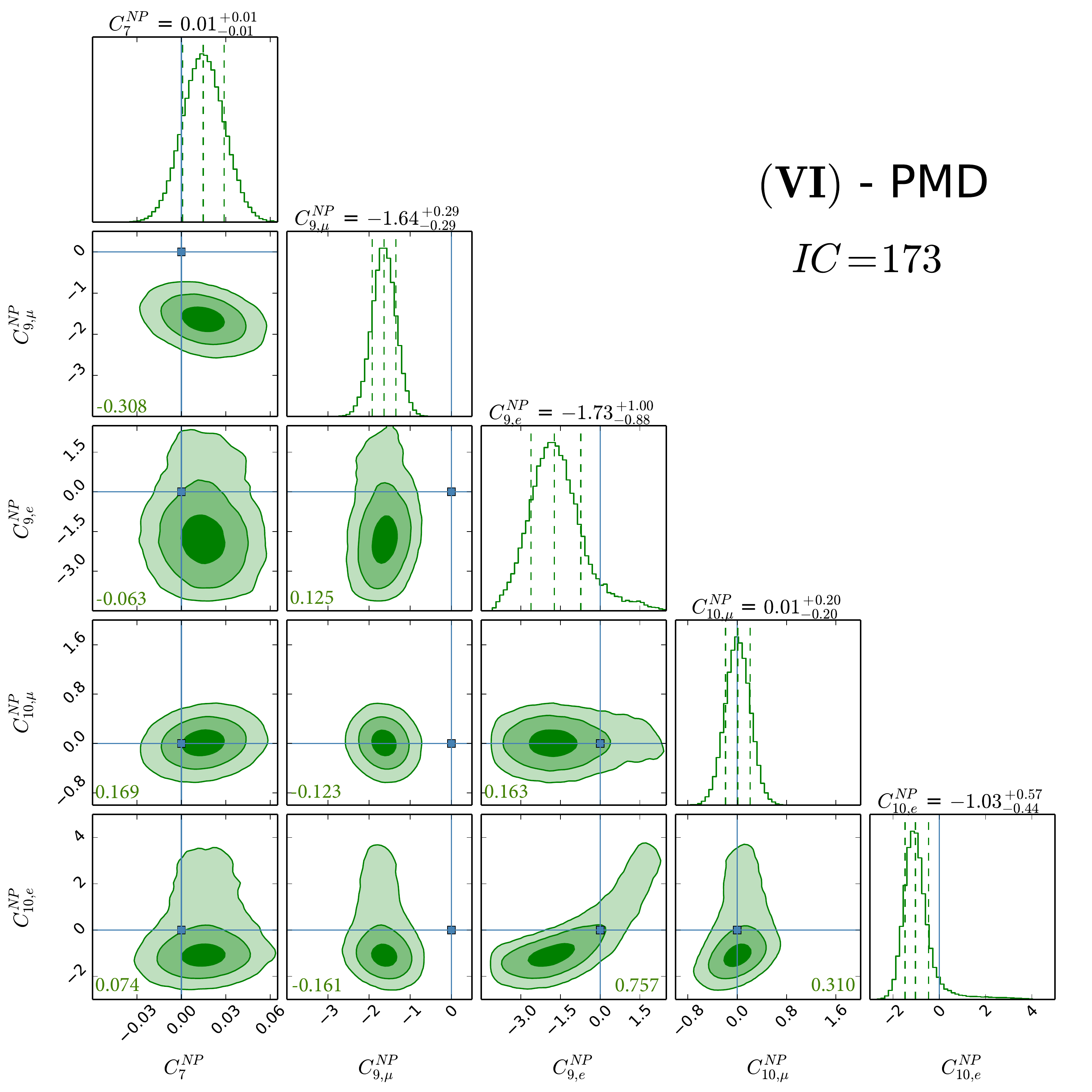}}
\subfigure{\includegraphics[width=.5\textwidth]{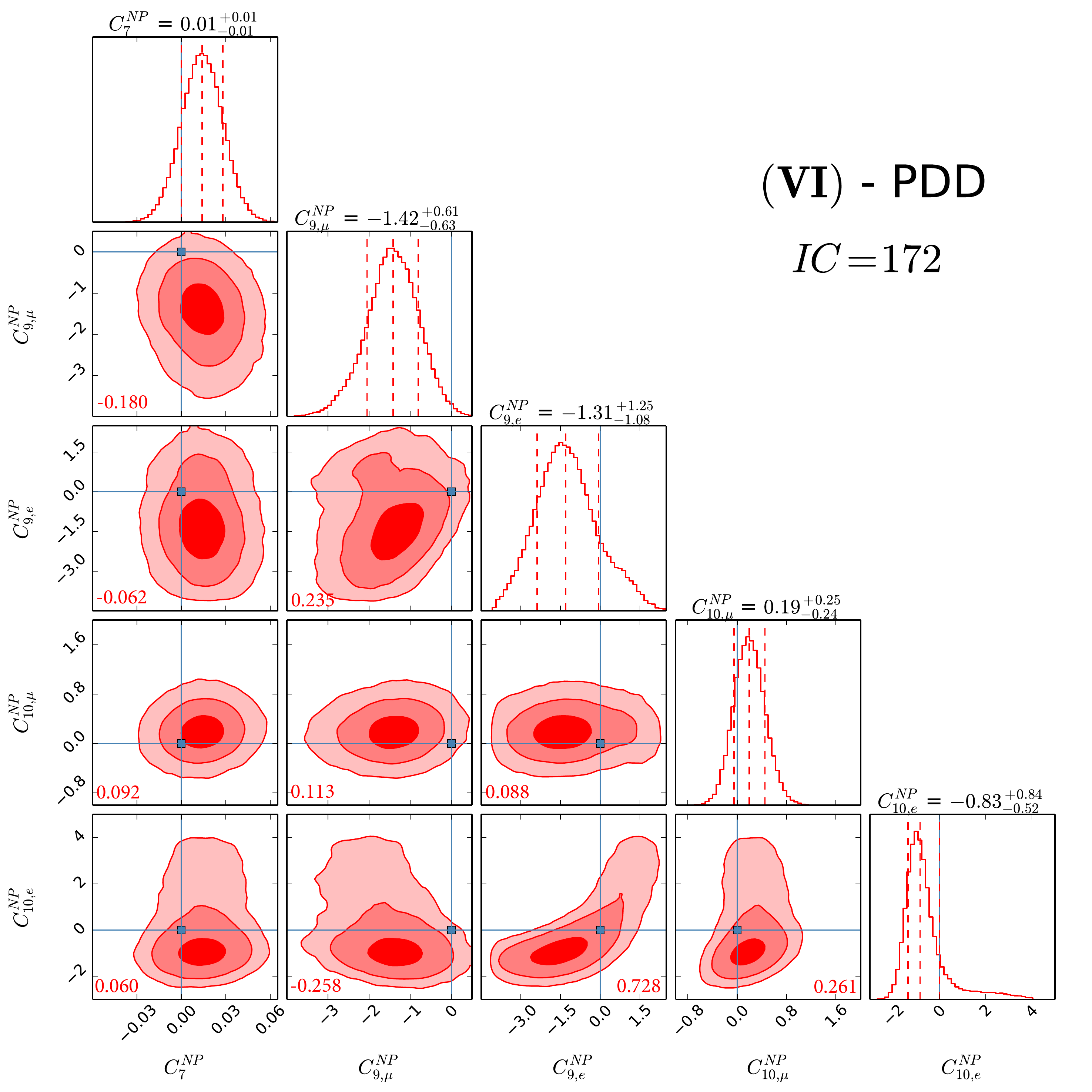}}
\caption{\textit{The five NP parameter fit using $C^{NP}_{7}$,
    $C^{NP}_{9,\mu}$, $C^{NP}_{9,e}$,
    $C^{NP}_{10,\mu}$ and $C^{NP}_{10,e}$. See caption of figure
    \ref{fig:fig1} for the colour coding and further details.}}
\label{fig:fig6}
\end{figure}

The results for NP WCs for the several cases that we study can be
found in figures~\ref{fig:fig1}--\ref{fig:fig6}, where the $IC$ value
for each model is also reported, and in
tables~\ref{tab:PMDpars}--\ref{tab:PDDpars} in \ref{sec:tab}.
In tables~\ref{tab:PMDobs}--\ref{tab:PDDobs}, we report the results of
the fit for observables of interest.  We observe that all cases have
comparable $IC$ values except cases \textbf{(IV)} and \textbf{(V)},
which are disfavoured in the PMD approach while they remain viable in
the PDD one.  The main difference between the two approaches is that
angular observables, in particular $P_5'$, call for NP in
$C_{9,\mu}^{NP}$ in the PMD approach, while they can be accommodated
within the SM in the PDD one.

Let us discuss the various cases in more detail.  It is important to
stress that the evidence of NP in our fit for cases
\textbf{(I)}--\textbf{(V)} is always larger than $3\sigma$ for one of
the semileptonic NP WCs used in the analysis, given the need of a
source of LFUV primarily from $R_{K,K^{*}}$ measurements. In
particular, we remark that in the PMD scenarios of cases \textbf{(I)}
and \textbf{(II)} we get evidence for NP at more than
$5\sigma$. However, looking at the corresponding PDD scenarios, the NP
evidence gets significantly reduced, roughly between $3\sigma$ and
$4\sigma$. The reduction in the significance comes from the larger
hadronic uncertainties in the PDD approach which weaken the
constraining power of the angular observables on the NP WCs.

Concerning case \textbf{(III)}, we observe very similar findings to the
ones obtained for case \textbf{(I)}, since the effective
coupling for the radiative dipole operator is well constrained,
especially from the inclusive $B \to X_s \gamma$ branching fraction.

Regarding case \textbf{(IV)}, in which we vary
the three NP parameters $C_{7}^{NP}, C^{NP}_{10,\mu}$ and
$C^{NP}_{10,e}$, the model comparison between
the PDD and PMD realization of this NP benchmark is  quite
informative: NP effects in the dipole operator and in the axial
semileptonic currents cannot address at the same time $R_{K,K^{*}}$
ratios and the $P_{5}'$ anomaly in a satisfactory way when we stick to
small non-factorizable QCD power corrections; however, this is no
longer true when we allow for a more conservative estimate of the
hadronic uncertainties. In particular, the tension in the
fit coming from the angular analysis of $B \to K^{*} \mu \mu $ can be
now addressed by large QCD effects as those given in
eq.~(\ref{eq:hlambda}), while a $C^{NP}_{10,e} \neq 0 $ at about $3\sigma$
can successfully describe all the observational hints of LFUV showed
by current measurements.
This interesting possibility of \textit{axial lepton-flavor violating NP} is not found in other
global analyses~\cite{Altmannshofer:2014rta,Descotes-Genon:2015uva,Chobanova:2017ghn,Altmannshofer:2017fio}, as it proceeds from the conservative treatment of hadronic uncertainties we proposed
in ref.~\cite{Ciuchini:2015qxb}.

Concerning tables~\ref{tab:PMDobs}--\ref{tab:PDDobs} of
~\ref{sec:tab}, we would like
to point out the pattern displayed by the
transverse ratios $R_{K^*}^T$ and $R_{\phi}^T$: cases \textbf{(I)}--\textbf{(III)}
predict these values to be $\sim 1$ with a small error, while
the remaining cases give different predictions with the central value
 ranging between $\sim 0.7$ and $\sim 0.8$. Therefore,
obtaining experimental information on transverse ratios may help
in discerning between the different NP scenarios.

We then show results for case \textbf{(V)}, in which we vary
$C_{7}^{NP}$, $C_{9,\mu}^{NP}$, $C_{9,e}^{NP}$ and correlate the
semileptonic vector and axial currents according to
$C_{9,\mu}^{NP}=-C_{10,\mu}^{NP}$ and $C_{9,e}^{NP}=-C_{10,e}^{NP}$.
In analogy to case \textbf{(IV)}, only within the PDD approach we find
for this NP benchmark a fairly good description of data, with
$C_{9,\mu}^{NP} = -C_{10,\mu}^{NP}$ compatible with zero at
$\sim 2\sigma$. Again, we are presented with the case where deviations
in angular observables are addressed by large QCD power corrections,
while LFUV is driven by semielectronic operators.  Looking back at
tables~\ref{tab:PMDobs}--\ref{tab:PDDobs}, we note that for this case,
as well as for case \textbf{(IV)} and \textbf{(VI)}, both transverse
and longitudinal muon over electron ratios in the central-$q^{2}$ bin,
namely $R^T_{K,K^{*},\phi} \,$ and $R^L_{K,K^{*},\phi} \,$, are
characterized by similar central values.

We close our presentation with an analysis of case \textbf{(VI)} in
which we float simultaneously $C_{7}^{NP}$, $C_{9,\mu}^{NP}$,
$C_{9,e}^{NP}$, $C_{10,\mu}^{NP}$, and $C_{10,e}^{NP}$. As can be seen
from figure~\ref{fig:fig6}, current measurements are informative
enough to constrain, at the same time, all the NP WCs both in the PMD
and PDD approaches.  In particular, within the latter case, a
nontrivial interplay among NP effects encoded both in
$C_{9,\mu }^{NP}$ and $C_{10,e }^{NP}$, together with the hadronic
contributions reported in table~\ref{tab:PDDpars}, produces the
weakest hint in favour of NP provided by our global analysis --
sitting between $2\sigma$ and $3\sigma$ level -- while allowing for a
very good description of the entire data set, similar to the other
cases.  The power corrections we found are larger than those obtained
in ref.~\cite{Khodjamirian:2010vf}, but smaller than those required by
the SM fit of $B\to K^*\mu\mu$~\cite{Ciuchini:2015qxb}. As discussed
in detail in refs.~\cite{Ciuchini:2016weo,Ciuchini:2017gva}, the size
obtained for the power corrections is compatible with the naive power
counting relative to the leading amplitude.  We stress (once again)
that a more optimistic attitude towards the estimate of QCD power
corrections (PMD approach) leads to the a much stronger claim in
favour of NP, at a statistical significance larger than $5\sigma$.

In tables~\ref{tab:PMDpars}--\ref{tab:PDDpars} we report mean and
standard deviation for the NP WCs and absolute values of $h_{\lambda}$
for all the cases considered in the analysis. It is also relevant to
observe that, once we switch on NP effects through $C^{NP}_{9,\mu}$ in
order to attempt at simultaneously explaining observables such as
$R_{K,K^{*}}$ and $P_{5}'$ in the PDD approach we find values for
$|h_\lambda^{(1,2)}|$ compatible with zero at $\sim
1\sigma$. Conversely, if we set $C^{NP}_{9,\mu}=0$ then a nonvanishing
$|h_{-}^{(2)}|$ is needed to account for the angular observables, as
found in ref.~\cite{Ciuchini:2015qxb}, showing that one cannot
disentangle hadronic uncertainties and NP in $B\to K^*\mu\mu$ at
present.


\section{Discussion}
\label{sec:conclusions}

In this work, we critically examined several BSM scenarios in order to
possibly explain the growing pattern of $B$ anomalies, recently
enriched by the $R_{K^*}$ measurement performed by the LHCb
collaboration~\cite{LHCb_RKstar}. We carried out our analysis in an
effective field theory framework, describing the non-factorizable
power corrections by means of 16 free parameters in our fit along the
lines of ref.~\cite{Ciuchini:2015qxb}.

We performed all our fits using two different hadronic models. The
first approach, labelled PMD, relies completely on the
phenomenological model from ref.~\cite{Khodjamirian:2010vf} and
corresponds to the more widely used choice in the literature. The
second one, named PDD, imposes the result of
ref.~\cite{Khodjamirian:2010vf} only at $q^2\le1$,\footnote{This
  choice is motivated in ref.~\cite{Ciuchini:2015qxb}.} allowing the
data to drive the hadronic contributions in the higher invariant mass
region.

Regarding the NP contributions, we analyze six different benchmark
scenarios, differentiated by distinct choices of NP WCs employed in
the fits. Case \textbf{(I)} allows for $C^{NP}_{9,\mu}$ and
$C^{NP}_{9,e}$, while case \textbf{(II)} considers the scenario with
$C^{NP}_{9,\mu}$ and $C^{NP}_{10,\mu}$; case \textbf{(III)} studies NP
effects coming as $C^{NP}_{7}$, $C^{NP}_{9,\mu}$ and $C^{NP}_{9,e}$,
and case \textbf{(IV)} is the same as the latter but with $C^{NP}_{10}$
instead of $C^{NP}_{9}$; case \textbf{(V)} studies the possibility
described in the third case with
$C_{10,\mu}^{NP} = - C_{9,\mu}^{NP}$  and
$C_{10,e}^{NP} = - C_{9,e}^{NP}$ enforced; finally, case
\textbf{(VI)} considers the general case with
all the five NP WCs being allowed to float independently. Our main results are
collected in figures~\ref{fig:fig1}--\ref{fig:fig6} and
also reported in tables~\ref{tab:PMDpars}--\ref{tab:PDDobs}.

The comparison of different scenarios using the $IC$ shows that all
the considered cases are on the same footing except for cases
\textbf{(IV)} and \textbf{(V)}. These cases are strongly disfavoured
in the PMD approach, as there is no $C_{9,\mu}^{NP}$ in case
\textbf{(IV)} to account for the deviation in $P_5'$, while
$C_{9,\mu}^{NP}$ is constrained by its correlation with
$C_{10,\mu}^{NP}$ and the measured value of BR$(B_s\to\mu\mu)$ in case
\textbf{(V)}.

In fact, from our analysis of radiative and (semi)leptonic
$B$ decays we identify two classes of viable NP scenarios:
\begin{itemize}
\item The widely studied $C^{NP}_{9,\mu} \neq 0$ scenario: from
  figures~\ref{fig:fig1}--\ref{fig:fig3}, we find a remarkable
  $\gtrsim 5\sigma$ evidence in favour of $C^{NP}_{9,\mu} \neq 0$ in
  the PMD approach. It is indeed nontrivial that a single NP WC can
  explain all the present anomalies in $b\to s$ transitions
  \cite{Beaujean:2013soa,Hurth:2013ssa,Altmannshofer:2014rta,Descotes-Genon:2015uva,Chobanova:2017ghn,Altmannshofer:2017fio}.
  However, in the more conservative PDD approach, the significance of
  a nonvanishing $C^{NP}_{9,\mu}$ drops to about $3\sigma$, mainly
  driven by LFUV.
\item An alternative scenario with nonvanishing $C^{NP}_{10,e}$, which
  emerges in the presence of large hadronic corrections to the
  infinite mass limit, namely our PDD approach. To our knowledge, a NP
  electronic axial current has not been studied in the literature,
  since it does not provide a satisfactory description of the angular
  observables within the commonly used PMD approach. We think that the
  present theoretical status of power correction calculations is not
  robust enough to discard this interesting NP scenario.
\end{itemize}

Finally the most general fit we performed, namely case \textbf{(VI)},
confirms in the PDD approach that both scenarios above are viable,
although a slight preference for $C^{NP}_{9,\mu} \neq 0$ is found.
More data are needed to assess what kind of NP scenario (if the
anomalies persist) is realized in Nature.

\begin{acknowledgements}
The research leading to these results has received
funding from the European Research Council under the European Union's
Seventh Framework Programme (FP/2007-2013)/ERC Grant Agreements
n. 279972 ``NPFlavour'' and n. 267985 ``DaMeSyFla''.
M.C. is associated to the Dipartimento di Matematica e
Fisica, Universit{\`a} di Roma Tre, and E.F. and L.S. are associated
to the Dipartimento di Fisica, Universit{\`a} di Roma ``La Sapienza''.

We wish to express our sincere gratitude to Alfredo Urbano for his
invaluable scientific correspondence concerning these flavourful
Easter eggs of New Physics.
\end{acknowledgements}

{\onecolumn
\appendix{
\begin{multicols}{2}
\section{Numerical Results}
\label{sec:tab}
In this appendix we present the tables with the most relevant
numerical results obtained from our global analysis. Mean and standard
deviation for the NP WCs and $h_{\lambda}$ absolute values are
reported in table~\ref{tab:PMDpars} for the PMD approach and in
table~\ref{tab:PDDpars} for the PDD one.\footnote{Percentiles for the
  NP WCs are reported in figures \ref{fig:fig1}-\ref{fig:fig6}.} In
table~\ref{tab:PMDobs} we list the results in the PMD approach
obtained for the key observables in the six NP scenarios. Analogous
results for the PDD approach can be found in table~\ref{tab:PDDobs}.
\end{multicols}
\vspace{1cm}
\begin{table*}[h!]
\centering{
\begin{tabular}{|c|c|c|c|c|c|c|}
\hline
&&&&&&\\[-2mm]
\textbf{Par.} & \textbf{(I)} & \textbf{(II)} & \textbf{(III)} & \textbf{(IV)} & \textbf{(V)} & \textbf{(VI)} \\[1mm]
\hline
&&&&&&\\[-2mm]
 $ C_{7}^{\rm NP} $
&
\phantom{-}
$-$
&
\phantom{-}
$-$
&
\phantom{-}
$0.015 \pm 0.014$
&
$-0.011 \pm 0.013$
&
\phantom{-}
$0.003 \pm 0.013$
&
\phantom{-}
$0.015 \pm 0.014$
\\
 $ C_{9,\mu}^{\rm NP}$
&
$-1.58 \pm 0.28$
&
$-1.53 \pm 0.25$
&
$-1.66 \pm 0.29$
&
\phantom{-}
$-$
&
$-0.54 \pm 0.17$
&
$-1.64 \pm 0.29$
\\
 $ C_{9,e}^{\rm NP}$
&
$-0.10 \pm 0.45$
&
\phantom{-}
$-$
&
$-0.18 \pm 0.46$
&
\phantom{-}
$-$
&
\phantom{-}
$0.09 \pm 0.25$
&
$-1.6 \pm 1.0$
\\
 $ C_{10,\mu}^{\rm NP}$
&
\phantom{-}
$-$
&
\phantom{-}
$0.03 \pm 0.16$
&
\phantom{-}
$-$
&
$-0.12 \pm 0.22$
&
\phantom{-}
$0.54 \pm 0.17$
&
\phantom{-}
$0.009 \pm 0.200$
\\
 $ C_{10,e}^{\rm NP}$
&
\phantom{-}
$-$
&
\phantom{-}
$-$
&
\phantom{-}
$-$
&
$-1.22 \pm 0.37$
&
$-0.09 \pm 0.25$
&
$-0.91 \pm 0.76$
\\

\hline
&&&&&&\\[-2mm]
 $ |h_0^{(0)}| \cdot 10^4$
& $2.1 \pm 1.2$
& $2.0 \pm 1.2$
& $2.2 \pm 1.3$
& $1.8 \pm 1.2$
& $1.3 \pm 1.0$
& $2.0 \pm 1.3$
\\
 $ |h_+^{(0)}| \cdot 10^4$
& $0.079 \pm 0.067$
& $0.079 \pm 0.067$
& $0.076 \pm 0.065$
& $0.083 \pm 0.069$
& $0.086 \pm 0.072$
& $0.076 \pm 0.064$
\\
 $ |h_-^{(0)}| \cdot 10^4$
& $0.53 \pm 0.19$
& $0.54 \pm 0.19$
& $0.52 \pm 0.19$
& $0.56 \pm 0.20$
& $0.60 \pm 0.21$
& $0.52 \pm 0.19$
\\

\hline
&&&&&&\\[-2mm]
 $ |h_0^{(1)}| \cdot 10^4$
& $0.30 \pm 0.23$
& $0.30 \pm 0.22$
& $0.30 \pm 0.23$
& $0.45 \pm 0.26$
& $0.32 \pm 0.24$
& $0.28 \pm 0.22$
\\
 $ |h_+^{(1)}| \cdot 10^4$
& $0.22 \pm 0.20$
& $0.22 \pm 0.19$
& $0.22 \pm 0.19$
& $0.21 \pm 0.19$
& $0.26 \pm 0.22$
& $0.22 \pm 0.19$
\\
 $ |h_-^{(1)}| \cdot 10^4$
& $0.23 \pm 0.19$
& $0.23 \pm 0.19$
& $0.23 \pm 0.20$
& $0.30 \pm 0.21$
& $0.32 \pm 0.22$
& $0.23 \pm 0.19$
\\

\hline
&&&&&&\\[-2mm]
 $ |h_+^{(2)}| \cdot 10^4$
& $0.052 \pm 0.045$
& $0.053 \pm 0.045$
& $0.052 \pm 0.044$
& $0.046 \pm 0.042$
& $0.064 \pm 0.053$
& $0.050 \pm 0.044$
\\
 $ |h_-^{(2)}| \cdot 10^4$
& $0.046 \pm 0.038$
& $0.046 \pm 0.039$
& $0.046 \pm 0.039$
& $0.092 \pm 0.050$
& $0.070 \pm 0.047$
& $0.045 \pm 0.038$
\\
\hline
\end{tabular}
}
  \caption{\it Results from the fit for WCs and hadronic contributions in the PMD approach. See Sec.~\ref{sec:eft&NP} for
  details on the six NP scenarios.}
  \label{tab:PMDpars}
\end{table*}

\begin{table*}[h!]
\centering{
\begin{tabular}{|c|c|c|c|c|c|c|}
\hline
&&&&&&\\[-1mm]
\textbf{Par.} & \textbf{(I)} & \textbf{(II)} & \textbf{(III)} & \textbf{(IV)} & \textbf{(V)} & \textbf{(VI)} \\[1mm]
\hline
&&&&&&\\[-2mm]
 $ C_{7}^{\rm NP} $
&
\phantom{-}
$-$
&
\phantom{-}
$-$
&
\phantom{-}
$0.013 \pm 0.014$
&
\phantom{-}
$0.008 \pm 0.014$
&
\phantom{-}
$0.011 \pm 0.014$
&
\phantom{-}
$0.014 \pm 0.014$
\\
 $ C_{9,\mu}^{\rm NP}$
&
$-1.47 \pm 0.63$
&
$-1.17 \pm 0.46$
&
$-1.58 \pm 0.64$
&
\phantom{-}
$-$
&
$-0.43 \pm 0.23$
&
$-1.43 \pm 0.64$
\\
 $ C_{9,e}^{\rm NP}$
&
\phantom{-}
$0.007 \pm 0.620$
&
\phantom{-}
$-$
&
$-0.08 \pm 0.63$
&
\phantom{-}
$-$
&
\phantom{-}
$0.21 \pm 0.29$
&
$-1.2 \pm 1.2$
\\
 $ C_{10,\mu}^{\rm NP}$
&
\phantom{-}
$-$
&
\phantom{-}
$0.26 \pm 0.23$
&
\phantom{-}
$-$
&
\phantom{-}
$0.27 \pm 0.26$
&
\phantom{-}
$0.43 \pm 0.23$
&
\phantom{-}
$0.20 \pm 0.25$
\\
 $ C_{10,e}^{\rm NP}$
&
\phantom{-}
$-$
&
\phantom{-}
$-$
&
\phantom{-}
$-$
&
$-0.86 \pm 0.4$
&
$-0.21 \pm 0.29$
&
$-0.60 \pm 0.99$
\\

\hline
&&&&&&\\[-2mm]
 $ |h_0^{(0)}| \cdot 10^4$
& $2.6 \pm 1.6$
& $2.3 \pm 1.4$
& $2.6 \pm 1.6$
& $1.7 \pm 1.3$
& $1.7 \pm 1.3$
& $2.6 \pm 1.6$
\\
 $ |h_+^{(0)}| \cdot 10^4$
& $0.075 \pm 0.066$
& $0.081 \pm 0.070$
& $0.077 \pm 0.067$
& $0.086 \pm 0.075$
& $0.087 \pm 0.075$
& $0.077 \pm 0.067$
\\
 $ |h_-^{(0)}| \cdot 10^4$
& $0.52 \pm 0.21$
& $0.55 \pm 0.22$
& $0.52 \pm 0.21$
& $0.60 \pm 0.23$
& $0.59 \pm 0.23$
& $0.53 \pm 0.21$
\\

\hline
&&&&&&\\[-2mm]
 $ |h_0^{(1)}| \cdot 10^4$
& $0.40 \pm 0.32$
& $0.41 \pm 0.34$
& $0.39 \pm 0.32$
& $0.50 \pm 0.36$
& $0.46 \pm 0.37$
& $0.40 \pm 0.33$
\\
 $ |h_+^{(1)}| \cdot 10^4$
& $0.40 \pm 0.29$
& $0.42 \pm 0.30$
& $0.40 \pm 0.29$
& $0.39 \pm 0.29$
& $0.42 \pm 0.30$
& $0.41 \pm 0.30$
\\
 $ |h_-^{(1)}| \cdot 10^4$
& $0.47 \pm 0.35$
& $0.52 \pm 0.38$
& $0.48 \pm 0.36$
& $0.82 \pm 0.46$
& $0.73 \pm 0.43$
& $0.50 \pm 0.37$
\\

\hline
&&&&&&\\[-2mm]
 $ |h_+^{(2)}| \cdot 10^4$
& $0.138 \pm 0.087$
& $0.160 \pm 0.099$
& $0.131 \pm 0.086$
& $0.139 \pm 0.094$
& $0.160 \pm 0.100$
& $0.145 \pm 0.095$
\\
 $ |h_-^{(2)}| \cdot 10^4$
& $0.112 \pm 0.085$
& $0.126 \pm 0.098$
& $0.111 \pm 0.083$
& $0.190 \pm 0.100$
& $0.170 \pm 0.110$
& $0.124 \pm 0.094$
\\
\hline
\end{tabular}
}
\caption{\it Results from the fit for WCs and hadronic contributions in the PDD approach. See Sec.~\ref{sec:eft&NP} for
  details on the six NP scenarios.}
  \label{tab:PDDpars}
 \end{table*}

\begin{table*}[h!]
\fontsize{6.3}{13}\selectfont
\centering{
\begin{tabular}{|c|c|c|c|c|c|c|c|}
\hline
&&&&&&&\\[-4mm]
\textbf{Obs.} & \textbf{Exp. value} & \textbf{(I)} & \textbf{(II)} & \textbf{(III)} & \textbf{(IV)} & \textbf{(V)} & \textbf{(VI)} \\[1mm]
\hline
&&&&&&&\\[-4mm]
 $ {R_K}_{[1,6]} $
& $0.753 \pm 0.090$
& $0.722 \pm 0.067$
& $0.703 \pm 0.047$
& $0.722 \pm 0.067$
& $0.781 \pm 0.055$
& $0.740 \pm 0.061$
& $0.724 \pm 0.067$
\\
 $ {R_{K^*}}_{[0.045,1.1]}$
& $0.680 \pm 0.093$
& $0.885 \pm 0.016$
& $0.881 \pm 0.016$
& $0.885 \pm 0.016$
& $0.839 \pm 0.024$
& $0.858 \pm 0.019$
& $0.843 \pm 0.030$
\\
 $ {R_{K^*}}_{[1.1,6]}$
& $0.707 \pm 0.102$
& $0.803 \pm 0.057$
& $0.786 \pm 0.049$
& $0.802 \pm 0.057$
& $0.713 \pm 0.065$
& $0.740 \pm 0.060$
& $0.717 \pm 0.067$
\\

\hline
&&&&&&&\\[-4mm]
 $ {R_{K^*}^L}_{[1.1,6]} $
& $-$
& $0.724 \pm 0.067$
& $0.705 \pm 0.049$
& $0.723 \pm 0.067$
& $0.722 \pm 0.065$
& $0.722 \pm 0.063$
& $0.723 \pm 0.068$
\\
 $ {R_{K^*}^T}_{[1.1,6]}$
& $-$
& $1.053 \pm 0.030$
& $1.046 \pm 0.069$
& $1.050 \pm 0.030$
& $0.692 \pm 0.067$
& $0.794 \pm 0.053$
& $0.75 \pm 0.25$
\\
 $ {R_{\phi}}_{[1.1,6]}$
& $-$
& $0.782 \pm 0.060$
& $0.764 \pm 0.048$
& $0.781 \pm 0.060$
& $0.720 \pm 0.064$
& $0.737 \pm 0.061$
& $0.717 \pm 0.062$
\\
 $ {R_{\phi}^L}_{[1.1,6]}$
& $-$
& $0.723 \pm 0.067$
& $0.704 \pm 0.049$
& $0.722 \pm 0.067$
& $0.728 \pm 0.064$
& $0.723 \pm 0.063$
& $0.721 \pm 0.067$
\\
 $ {R_{\phi}^T}_{[1.1,6]}$
& $-$
& $1.044 \pm 0.032$
& $1.038 \pm 0.066$
& $1.043 \pm 0.033$
& $0.691 \pm 0.067$
& $0.794 \pm 0.053$
& $0.75 \pm 0.24$
\\

\hline
&&&&&&&\\[-4mm]
 $ {P_5}_{[4,6]}^{\mbox{\tiny LHCb}} $
& $-0.301 \pm 0.160$
& $-0.423 \pm 0.060$
& $-0.427 \pm 0.061$
& $-0.420 \pm 0.059$
& $-0.556 \pm 0.063$
& $-0.587 \pm 0.053$
& $-0.431 \pm 0.061$
\\
 $ {P_5}_{[6,8]}^{\mbox{\tiny LHCb}}$
& $-0.505 \pm 0.124$
& $-0.602 \pm 0.059$
& $-0.607 \pm 0.061$
& $-0.598 \pm 0.059$
& $-0.677 \pm 0.066$
& $-0.704 \pm 0.057$
& $-0.61 \pm 0.06$
\\
 $ {P_5}_{[4,6]}^{\mbox{\tiny ATLAS}}$
& $-0.26 \pm 0.39$
& $-0.423 \pm 0.060$
& $-0.427 \pm 0.061$
& $-0.420 \pm 0.059$
& $-0.556 \pm 0.063$
& $-0.587 \pm 0.053$
& $-0.431 \pm 0.061$
\\
 $ {P_5}_{[4.3,6]}^{\mbox{\tiny CMS}}$
& $-0.955 \pm 0.268$
& $-0.442 \pm 0.059$
& $-0.447 \pm 0.061$
& $-0.440 \pm 0.059$
& $-0.572 \pm 0.063$
& $-0.603 \pm 0.053$
& $-0.451 \pm 0.061$
\\
 $ {P_5}_{[6,8.68]}^{\mbox{\tiny CMS}}$
& $-0.660 \pm 0.220$
& $-0.623 \pm 0.060$
& $-0.626 \pm 0.061$
& $-0.618 \pm 0.059$
& $-0.688 \pm 0.067$
& $-0.711 \pm 0.058$
& $-0.63 \pm 0.06$
\\

\hline
&&&&&&&\\[-4mm]
 $ {P_5}_{[4,8]}^{\mbox{\tiny Belle}} $
& $-0.025 \pm 0.318$
& $-0.520 \pm 0.057$
& $-0.524 \pm 0.059$
& $-0.516 \pm 0.057$
& $-0.620 \pm 0.063$
& $-0.649 \pm 0.053$
& $-0.528 \pm 0.059$
\\
 $ {P_{5,e}}_{[4,8]}^{\mbox{\tiny Belle}}$
& $-0.510 \pm 0.272$
& $-0.782 \pm 0.054$
& $-0.794 \pm 0.039$
& $-0.782 \pm 0.054$
& $-0.536 \pm 0.063$
& $-0.710 \pm 0.044$
& $-0.42 \pm 0.23$
\\
\hline
\end{tabular}
}
  \caption{\it Experimental results (with symmetrized errors) and results from the
  fit for key observables in the PMD approach. See Sec.~\ref{sec:eft&NP} for
  details on the six NP scenarios.}
  \label{tab:PMDobs}
\end{table*}

\begin{table*}[h!]
\fontsize{6.3}{13}\selectfont
\begin{tabular}{|c|c|c|c|c|c|c|c|}
\hline
&&&&&&&\\[-4mm]
\textbf{Obs.} & \textbf{Exp. value} & \textbf{(I)} & \textbf{(II)} & \textbf{(III)} & \textbf{(IV)} & \textbf{(V)} & \textbf{(VI)} \\[1mm]
\hline
&&&&&&&\\[-4mm]
 $ {R_K}_{[1,6]} $
& $0.753 \pm 0.090$
& $0.724 \pm 0.068$
& $0.714 \pm 0.064$
& $0.723 \pm 0.067$
& $0.775 \pm 0.057$
& $0.740 \pm 0.062$
& $0.719 \pm 0.073$
\\
 $ {R_{K^*}}_{[0.045,1.1]}$
& $0.680 \pm 0.093$
& $0.883 \pm 0.017$
& $0.873 \pm 0.016$
& $0.883 \pm 0.017$
& $0.842 \pm 0.024$
& $0.861 \pm 0.019$
& $0.847 \pm 0.030$
\\
 $ {R_{K^*}}_{[1.1,6]}$
& $0.707 \pm 0.102$
& $0.803 \pm 0.059$
& $0.777 \pm 0.053$
& $0.801 \pm 0.059$
& $0.714 \pm 0.066$
& $0.755 \pm 0.059$
& $0.722 \pm 0.069$
\\

\hline
&&&&&&&\\[-4mm]
 $ {R_{K^*}^L}_{[1.1,6]} $
& $-$
& $0.726 \pm 0.070$
& $0.719 \pm 0.061$
& $0.723 \pm 0.070$
& $0.726 \pm 0.065$
& $0.729 \pm 0.063$
& $0.719 \pm 0.073$
\\
 $ {R_{K^*}^T}_{[1.1,6]}$
& $-$
& $1.050 \pm 0.033$
& $0.959 \pm 0.083$
& $1.046 \pm 0.034$
& $0.688 \pm 0.069$
& $0.824 \pm 0.049$
& $0.8 \pm 0.32$
\\
 $ {R_{\phi}}_{[1.1,6]}$
& $-$
& $0.782 \pm 0.062$
& $0.761 \pm 0.054$
& $0.779 \pm 0.062$
& $0.722 \pm 0.065$
& $0.749 \pm 0.060$
& $0.718 \pm 0.063$
\\
 $ {R_{\phi}^L}_{[1.1,6]}$
& $-$
& $0.724 \pm 0.070$
& $0.716 \pm 0.061$
& $0.721 \pm 0.070$
& $0.731 \pm 0.064$
& $0.728 \pm 0.063$
& $0.716 \pm 0.071$
\\
 $ {R_{\phi}^T}_{[1.1,6]}$
& $-$
& $1.039 \pm 0.036$
& $0.955 \pm 0.079$
& $1.036 \pm 0.036$
& $0.693 \pm 0.068$
& $0.828 \pm 0.049$
& $0.79 \pm 0.3$
\\

\hline
&&&&&&&\\[-4mm]
 $ {P_5}_{[4,6]}^{\mbox{\tiny LHCb}} $
& $-0.301 \pm 0.160$
& $-0.398 \pm 0.066$
& $-0.405 \pm 0.068$
& $-0.401 \pm 0.066$
& $-0.435 \pm 0.065$
& $-0.421 \pm 0.068$
& $-0.417 \pm 0.068$
\\
 $ {P_5}_{[6,8]}^{\mbox{\tiny LHCb}}$
& $-0.505 \pm 0.124$
& $-0.536 \pm 0.080$
& $-0.521 \pm 0.083$
& $-0.539 \pm 0.080$
& $-0.521 \pm 0.080$
& $-0.508 \pm 0.082$
& $-0.545 \pm 0.083$
\\
 $ {P_5}_{[4,6]}^{\mbox{\tiny ATLAS}}$
& $-0.26 \pm 0.39$
& $-0.398 \pm 0.066$
& $-0.405 \pm 0.068$
& $-0.401 \pm 0.066$
& $-0.435 \pm 0.065$
& $-0.421 \pm 0.068$
& $-0.417 \pm 0.068$
\\
 $ {P_5}_{[4.3,6]}^{\mbox{\tiny CMS}}$
& $-0.955 \pm 0.268$
& $-0.414 \pm 0.067$
& $-0.420 \pm 0.068$
& $-0.417 \pm 0.067$
& $-0.447 \pm 0.066$
& $-0.434 \pm 0.068$
& $-0.433 \pm 0.068$
\\
 $ {P_5}_{[6,8.68]}^{\mbox{\tiny CMS}}$
& $-0.660 \pm 0.220$
& $-0.550 \pm 0.083$
& $-0.531 \pm 0.086$
& $-0.553 \pm 0.083$
& $-0.529 \pm 0.084$
& $-0.514 \pm 0.085$
& $-0.556 \pm 0.086$
\\

\hline
&&&&&&&\\[-4mm]
 $ {P_5}_{[4,8]}^{\mbox{\tiny Belle}} $
& $-0.025 \pm 0.318$
& $-0.472 \pm 0.070$
& $-0.467 \pm 0.072$
& $-0.475 \pm 0.070$
& $-0.481 \pm 0.070$
& $-0.467 \pm 0.072$
& $-0.486 \pm 0.072$
\\
 $ {P_{5,e}}_{[4,8]}^{\mbox{\tiny Belle}}$
& $-0.510 \pm 0.272$
& $-0.725 \pm 0.075$
& $-0.664 \pm 0.095$
& $-0.733 \pm 0.075$
& $-0.424 \pm 0.063$
& $-0.570 \pm 0.066$
& $-0.41 \pm 0.22$
\\
\hline
\end{tabular}
  \caption{\it Experimental results (with symmetrized errors) and results from the
  fit for key observables in the PDD approach. See Sec.~\ref{sec:eft&NP} for
  details on the six NP scenarios.}
  \label{tab:PDDobs}
 \end{table*}
 }
 }

\begin{multicols}{2}
\bibliographystyle{JHEP}
\bibliography{btosll2017}
\end{multicols}
\end{document}